\documentclass[fleqn,usenatbib]{mnras}

\usepackage{newtxtext}

\usepackage{eso-pic}

\AddToShipoutPictureBG*{%
  \AtPageUpperLeft{%
    \hspace{0.75\paperwidth}%
    \raisebox{-3.5\baselineskip}{%
      \makebox[0pt][l]{\textnormal{DES 2014-0015}}
}}}%

\AddToShipoutPictureBG*{%
  \AtPageUpperLeft{%
    \hspace{0.75\paperwidth}%
    \raisebox{-4.5\baselineskip}{%
      \makebox[0pt][l]{\textnormal{FERMILAB-PUB-16-021-AE}}
}}}%

\usepackage[T1]{fontenc}
\usepackage{ae,aecompl}

\usepackage{graphicx}	
\usepackage{amsmath,amsfonts,amssymb}
\usepackage{makeidx}
\usepackage{epstopdf} 
\usepackage{booktabs,fixltx2e}
\usepackage[flushleft]{threeparttable}

\usepackage{multirow}
\usepackage{rotating}
\usepackage{color}

\newcommand{\Rtwohundred}{$R_{200}$}
\newcommand{\LMU}{$^{1}$}
\newcommand{\ECU}{$^{2}$}
\newcommand{\MPE}{$^{3}$}
\newcommand{\CTIO}{$^{4}$}
\newcommand{\UCL}{$^{5}$}
\newcommand{\Rhodes}{$^{6}$}
\newcommand{\Colby}{$^{7}$}
\newcommand{\Harvard}{$^{8}$}
\newcommand{\CNRS}{$^{9}$}
\newcommand{\Sorbonne}{$^{10}$}
\newcommand{\Carnegie}{$^{11}$}
\newcommand{\ICG}{$^{12}$}
\newcommand{\LINEA}{$^{13}$}
\newcommand{\ONB}{$^{14}$}
\newcommand{\UIA}{$^{15}$}
\newcommand{\NCSA}{$^{16}$}
\newcommand{\ICE}{$^{17}$}
\newcommand{\IFAE}{$^{18}$}
\newcommand{\Southhampton}{$^{19}$}
\newcommand{\FNAL}{$^{20}$}
\newcommand{\UPPA}{$^{21}$}
\newcommand{\JPL}{$^{22}$}
\newcommand{\UMA}{$^{23}$}
\newcommand{\UMP}{$^{24}$}
\newcommand{\KavliC}{$^{25}$}
\newcommand{\UFA}{$^{26}$}
\newcommand{\KavliS}{$^{27}$}
\newcommand{\SLAC}{$^{28}$}
\newcommand{\CAPP}{$^{29}$}
\newcommand{\OSUP}{$^{30}$}
\newcommand{\Montreal}{$^{31}$}
\newcommand{\AAO}{$^{32}$}
\newcommand{\Texas}{$^{33}$}
\newcommand{\OSUA}{$^{34}$}
\newcommand{\KavliM}{$^{35}$}
\newcommand{\Princeton}{$^{36}$}
\newcommand{\ICRE}{$^{37}$}
\newcommand{\UMelb}{$^{38}$}
\newcommand{\Sussex}{$^{39}$}
\newcommand{\UAP}{$^{40}$}
\newcommand{\CIE}{$^{41}$}
\newcommand{\IFB}{$^{42}$}
\newcommand{\UAH}{$^{43}$}
\newcommand{\CFA}{$^{44}$}
\newcommand{\UCD}{$^{45}$}
\newcommand{\ANL}{$^{46}$}

\title[Galaxy Populations in Massive Clusters to $z=1.1$]{Galaxy Populations in Massive Galaxy Clusters to $z=1.1$: Color Distribution, Concentration, Halo Occupation Number and Red Sequence Fraction}
\author[C. Hennig et al.]{
C.~Hennig\LMU$^,$\ECU,
J.~J.~Mohr\LMU$^,$\ECU$^,$\MPE,
A.~Zenteno\CTIO$^,$\LMU,
S.~Desai\LMU$^,$\ECU,
J.~P.~Dietrich\LMU$^,$\ECU,
S.~Bocquet\LMU$^,$\ECU,
\newauthor
V.~Strazzullo\LMU$^,$\ECU,
A.~Saro\LMU$^,$\ECU,
T.~M.~C.~Abbott\CTIO,
F.~B.~Abdalla\UCL$^,$\Rhodes,
M.~Bayliss\Colby$^,$\Harvard,
\newauthor
A.~Benoit-L{\'e}vy\CNRS$^,$\UCL$^,$\Sorbonne,
R.~A.~Bernstein\Carnegie,
E.~Bertin\CNRS$^,$\Sorbonne,
D.~Brooks\UCL,
R.~Capasso\LMU$^,$\ECU,
\newauthor
D.~Capozzi\ICG,
A.Carnero\LINEA$^,$\ONB,
M.~Carrasco~Kind\UIA$^,$\NCSA,
J.~Carretero\ICE$^,$\IFAE,
I.~Chiu\LMU$^,$\ECU,
\newauthor
C.~B.~D'Andrea\ICG$^,$\Southhampton,
L.~N.~daCosta\LINEA$^,$\ONB,
H.~T.~Diehl\FNAL,
P.~Doel\UCL,
T.~F.~Eifler\UPPA$^,$\JPL,
\newauthor
A.~E.~Evrard\UMA$^,$\UMP,
A.~Fausti-Neto\LINEA,
P.~Fosalba\ICE,
J.~Frieman\FNAL$^,$\KavliC,
C.~Gangkofner\LMU$^,$\ECU,
\newauthor
A.~Gonzalez\UFA,
D.~Gruen\KavliS$^,$\SLAC,
R.~A.~Gruendl\UIA$^,$\NCSA,
N.~Gupta\LMU$^,$\ECU$^,$\MPE,
G.~Gutierrez\FNAL,
\newauthor
K.~Honscheid\CAPP$^,$\OSUP,
J.~Hlavacek-Larrondo\Montreal,
D.~J.~James\CTIO,
K.~Kuehn\AAO,
N.~Kuropatkin\FNAL,
\newauthor
O.~Lahav\UCL,
M.~March\UPPA,
J.~L.~Marshall\Texas,
P.~Martini\CAPP$^,$\OSUA,
M.~McDonald\KavliM,
P.~Melchior\Princeton,
\newauthor
C.~J.~Miller\UMA$^,$\UMP,
R.~Miquel\ICRE$^,$\IFAE,
E.~Neilsen\FNAL,
B.~Nord\FNAL,
R.~Ogando\LINEA$^,$\ONB,
A.~A.~Plazas\JPL,
\newauthor
C.~Reichardt\UMelb,
A.~K.~Romer\Sussex,
E.~Rozo\UAP,
E.~S.~Rykoff\KavliS$^,$\SLAC,
E.~Sanchez\CIE,
\newauthor
B.~Santiago\IFB$^,$\LINEA,
M.~Schubnell\UMP,
I.~Sevilla-Noarbe\CIE,
R.~C.~Smith\CTIO,
M.~Soares-Santos\FNAL,
\newauthor
F.~Sobreira\LINEA,
B.~Stalder\UAH$^,$\CFA,
S.A.~Stanford\UCD,
E.~Suchyta\UPPA,
M.~E.~C.~Swanson\NCSA,
\newauthor
G.~Tarle\UMP,
D.~Thomas\ICG,
V.~Vikram\ANL,
A.~R.~Walker\CTIO,
Y.~Zhang\FNAL}

\date{Accepted XXX. Received YYY; in original form ZZZ}
\pubyear{2016}
\begin{document}

\label{firstpage}

\pagerange{\pageref{firstpage}--\pageref{lastpage}}

\maketitle

\begin{abstract}
We study the galaxy populations in 74 Sunyaev Zeldovich Effect (SZE) selected clusters from the South Pole Telescope (SPT) survey that have been imaged in the science verification phase of the Dark Energy Survey (DES). The sample extends up to $z\sim 1.1$ with $4 \times 10^{14} M_{\odot}\le M_{200}\le 3\times 10^{15} M_{\odot}$.  Using the band containing the 4000~\AA\ break and its redward neighbor, we study the color-magnitude distributions of cluster galaxies to $\sim m_*+2$, finding: (1) the intrinsic rest frame $g-r$ color width of the red sequence (RS) population is $\sim$0.03 out to $z\sim0.85$ with a preference for an increase to $\sim0.07$ at $z=1$ and (2) the prominence of the RS declines beyond $z\sim0.6$.  The spatial distribution of cluster galaxies is well described by the NFW profile out to $4R_{200}$ with a concentration of $c_{\mathrm{g}} = 3.59^{+0.20}_{-0.18}$, $5.37^{+0.27}_{-0.24}$ and $1.38^{+0.21}_{-0.19}$ for the full, the RS and the blue non-RS populations, respectively, but with $\sim40$\% to 55\% cluster to cluster variation and no statistically significant redshift or mass trends. The number of galaxies within the virial region $N_{200}$ exhibits a mass trend indicating that the number of galaxies per unit total mass is lower in the most massive clusters, and shows no significant redshift trend. The red sequence (RS) fraction within $R_{200}$ is $(68\pm3)$\% at $z=0.46$, varies from $\sim$55\% at $z=1$ to $\sim$80\% at $z=0.1$, and exhibits intrinsic variation among clusters of $\sim14$\%. We discuss a model that suggests the observed redshift trend in RS fraction favors a transformation timescale for infalling field galaxies to become RS galaxies of 2 to 3~Gyr.    
\end{abstract}




\section{Introduction}

Galaxy clusters were first systematically cataloged based on optical observations \citep{abell58,zwicky61}. These clusters were primarily nearby systems, and they were mainly characterized by their richness, compactness and distance. Today, other techniques are widely used for detecting galaxy clusters.  One of the most widely used is based upon an observational signature that arises through the interaction of the hot intra-cluster medium with the low energy cosmic-microwave-background (CMB) photons. This so-called thermal Sunyaev Zel'dovich effect (SZE) is a spectral distortion of the CMB due to inverse-Compton scattering of CMB photons with the energetic galaxy cluster electrons \citep{sunyaev72}. The surface brightness of the SZE is independent of redshift and the integrated thermal SZE signature is expected to be tightly correlated with the cluster virial mass \citep[e.g.,][]{holder01b, andersson11}.  

Using the SZE for cluster selection allows us to identify high purity, approximately mass-limited cluster samples that span the full redshift range over which galaxy clusters exist \citep{song12b,bleem15}.  Together with multi-band optical data, these cluster samples enable studies of galaxy cluster properties, including the luminosity function of the cluster galaxies, the stellar mass fraction, the radial profile and the distribution of galaxy color \citep[e.g.,][]{mancone10,zenteno11,mancone12b,hilton13,chiu16a}. These measurements allow us to gain insights into galaxy formation and evolution and to assess the degree to which these processes are affected by environment and vary over cosmic time \citep[e.g.,][]{butcher84,stanford98,lin06,brodwin13,vdB15}.

A picture of the galaxy populations inside and outside clusters has emerged where the stellar mass functions (SMFs) for the passive and star forming galaxies are independent of environment, but the mix of these populations changes as one moves from the field to the cluster (\citealp[e.g.,][]{muzzin12}\citealp[; see also][]{binggeli88,jerjen97,andreon98}).  The redshift variation of the red fraction has been shown to provide constraints on the timescales on which infalling field galaxies are transformed to red sequence (RS) galaxies \citep{mcgee09}.  Moreover, the scatter of the RS and its variation with redshift has been used to constrain the variation in star formation histories and the timescale since the formation of the bulk of stars within RS galaxies \citep[e.g.,][]{aragon-salamanca93,stanford98,mei09,hilton09,papovich10}.  

The use of a homogeneously selected cluster sample extending over a broad redshift range, which has relatively uniform depth, multi-band imaging, enables one to carry out a systematic study of the population variations as a function of redshift and cluster mass.  This homogeneity allows us to compare high-z and low-z clusters without having to resort to combining our sample with those in the literature, an approach that is complicated by differences in analysis techniques and methods used to estimate cluster properties.  Moreover, the selection of the cluster sample using a method that is independent of the galaxy populations makes for a more straightforward interpretation of the observed properties of the cluster galaxies and their evolution.

In this paper, we report on our analysis of the color distribution and the radial profile of the cluster galaxy population through examination of the full population, the RS population and the non-RS or blue population.  Our primary goal is to understand how the cluster galaxy populations change with redshift and cluster mass.  
We focus on magnitudes and colors that are extracted from similar portions of the rest frame spectrum over the full redshift range.

Our sample arises from the overlap between the Dark Energy Survey \citep{DES05} science verification (DES-SV) data and the existing South Pole Telescope (SPT) 2500~deg$^2$ mm-wave survey  \citep[SPT-SZ; e.g.,][]{story13}.  The sample of SPT-SZ cluster candidates overlapping the DES-SV data  contains 74 clusters. Our analysis follows the optical study of the first four SZE selected clusters in \citet{zenteno11}, and is complementary to the analysis of the 26 most massive clusters extracted from the full SPT survey \citep{zenteno16}.  

This paper is organized as follows: Section~\ref{sec:data} describes the DES-SV observations and data reduction as well as the SPT selected cluster sample. In Section~\ref{sec:clusterproperties} we describe the cluster sample properties, presenting redshifts and masses.  In Section~\ref{sec:galaxypopulations} we present measurements of the radial and color distributions. We end with the conclusions in Section~\ref{sec:conclusions}. 

In this work, unless otherwise specified, we assume a flat $\Lambda$CDM Cosmology. The cluster masses refer to $M_{200,c}$, the mass enclosed within a virial sphere of radius \Rtwohundred, in which the mean matter density is equal to 200 times the critical density at the observed cluster redshift. The critical density is $\rho_{\mathrm{c}}(z) = 3H^2(z)/8\pi G$, where $H(z)$ is the Hubble parameter. We use the best fit cosmological parameters from \citet{bocquet15}: $\Omega_\mathrm{m}=0.292$ and $H_0=68.6$~km~s$^{-1}$Mpc$^{-1}$; these are derived through a combined analysis of the SPT cluster population, the WMAP constraints on the cosmic microwave background (CMB) anisotropy, supernovae distance measurements and baryon acoustic oscillation distance measurements.


\section{Observations and Data reduction}
\label{sec:data}

There are 74 clusters detected by SPT with signal to noise ratio $\xi>4.5$, which are imaged within the DES-SV data and have deep photometric coverage around the cluster position.  Below we describe how the sample of 74 multi-band coadds and associated calibrated galaxy catalogs are produced.

\subsection{DECam Data Processing and Calibration}
The DES-SV observations were acquired between November 1, 2012 and February 22, 2013 using the Dark Energy Camera \citep[DECam;][]{flaugher15}. The data were processed with an improved version of the pipeline used to process the Blanco Cosmology Survey Data \citep{desai12}, which has its heritage in the early DES data management system \citep{ngeow06,mohr08,mohr12}.  Following a data flow similar to that adopted for the BCS processing, we process data from every night using the single epoch pipeline.  The raw data from the telescope are first crosstalk corrected. For DECam, the crosstalk matrix includes negative co-efficients and also non-linear corrections for certain CCDs/amplifiers. Single-epoch images are then produced through a bias subtraction and dome flat correction. We implement a pixel scale correction to reduce the positional variation in the zeropoint or, equivalently, to flatten the zeropoint surface within the individual CCD detectors. No illumination or fringe corrections are applied; we adopt a star flat procedure to photometrically flatten the images.  In particular, we stack DES-SV stellar photometry from photometric observations in detector coordinates and determine for each band the persistent photometric residual in stellar photometry as a function of position.  We use this to create a position dependent photometric scale factor that further flattens the zeropoint surface within each detector and also brings all detectors to a common zeropoint \citep[see also][]{regnault09,schlafly12}.

First pass astrometric calibration is carried out exposure by exposure using the {\tt SCAMP} Astromatic software \citep{bertin06} and by calibrating to the 2MASS catalog \citep{skrutskie06}.  In this approach we use as input a high quality distortion map for the detector that we determine through a {\tt SCAMP} run of a large collection of overlapping exposures. The residual scatter of our first pass astrometry around 2MASS is approximately 200 milli-arcseconds, which is dominated by the 2MASS positional uncertainties.  In a second pass prior to the coaddition, we recalibrate the astrometry using {\tt SCAMP} and the full collection of overlapping DECam images around a particular area of interest on the sky (i.e. where there is a known SPT cluster).  This reduces the relative root mean square internal astrometric scatter around the best fit to 20~milli-arcseconds, which is a factor of a few improvement over the internal scatter in the first pass calibration.  For the data used in these analyses we find the first pass astrometric solution to be adequate for our needs.   

Cataloging is carried out using the model-fitting capabilities of SExtractor \citep{bertin96}, where we create position dependent point spread function (PSF) models for each image using PSFEx and then use these PSF models to evaluate a variety of customized, PSF corrected model magnitudes, object positions, morphology measures and star-galaxy classifiers.

Once all the data from each night are processed using the single epoch pipeline, we then photometrically calibrate the data and build coadds centered around each of the SPT cluster candidates.  We determine a relative photometric calibration using common stars within overlapping images.  We create median-combined coadd images using PSF homogenization to a common Moffat profile \citep{moffat69} with a full width at half maximum (FWHM) tuned to be the median of all the input single epoch images for each band. For coadd cataloging we first create a chi-square detection image \citep{szalay99} using the $i$ and $z$ band coadd images, and then we catalog in dual image mode, using a common detection image across all bands.  We use catalogs extracted from the PSF homogenized coadd images, because we have identified failure modes in the star-galaxy classification and in the centrally weighted galaxy colors in the non-homogenized coadds that are caused by discontinuities in the spatial variation of the PSF.

For absolute photometric  calibration of the final catalog we calibrate the color differences among different band combinations using the DECam stellar locus, where we first calibrate ($g-r$) vs ($r-i$) and then keeping the ($r-i$) offset fixed, we calibrate ($r-i$) vs ($i-z$). The absolute calibration  comes from the 2MASS $J$ band.  We do not use any $Y$ band data for this analysis.

\begin{figure}
\vskip-0.35cm
 \includegraphics [width=0.48\textwidth]{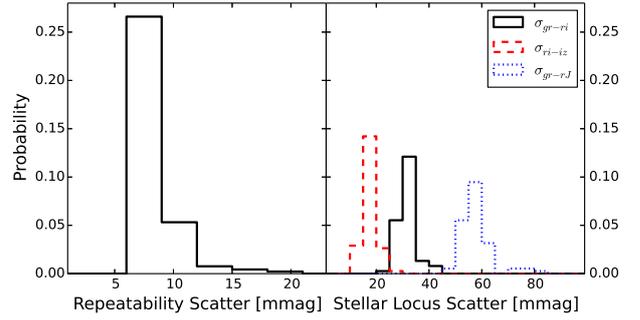}
 \vskip-0.2cm
 \caption [Scatter distributions in repeatability and stellar locus for DES]{Distribution of single epoch photometric repeatability scatter (left) for multiple measurements of the same stars in our ensemble of 74 clusters.  The bands $griz$ have a median scatter of 7.6, 7.6, 7.7 and 8.3~mmag, respectively.  The distribution of orthogonal scatter about the stellar locus is shown on the right for three color-color spaces.  The scatter distributions for these three spaces have a median scatter of 17, 32 and 57~mmag.
 \label{fig:catalogQA}}
 \vskip-0.25cm
\end{figure}

To determine the stellar locus in the DECam system we bootstrap from calibrated SDSS photometry.  We do this by determining the color terms between the DECam and SDSS systems using DECam observations of calibrated stars within the SDSS system.  The first order color terms we find are -0.088,-0.1079,-0.3080 and -0.0980 for $griz$ bands, respectively, where we use the color $g-r$ for $g$ and $r$ band and $r-i$ for $i$ and $z$.  With these color terms we then use calibrated SDSS photometry to predict the DECam stellar locus.  In this step we restrict our analysis to those stars with colors that lie in the range where the linear color correction is accurate at better than 1\%.  We then use this predicted stellar locus to calibrate the offsets in an empirical DECam stellar locus that we extract from selected high quality observations of a portion of the survey.

For each calibrated tile we evaluate the quality of the images and catalogs using the scatter around the stellar locus and the scatter obtained from photometric repeatability tests.  Figure~\ref{fig:catalogQA} (right) contains a plot of the orthogonal scatter of stars about the stellar locus in three different color-color spaces $r$-$i$ vs. $i$-$z$, $g$-$r$ vs. $r$-$i$ and $g$-$r$ vs. $r$-$J$.  The median scatter about the stellar locus in these three spaces is 17, 32 and 57~mmag, respectively.  These values are comparable to the stellar locus scatter 
in a recent PS1 analysis \citep{liu15} and better than values obtained from the BCS or SDSS datasets \citep{desai12}. 

In the photometric repeatability tests we compare the magnitude differences between multiple observations of the same object that are obtained from different single-epoch images which contribute to the coadd tile.  Figure~\ref{fig:catalogQA} (left) contains a plot of the distribution of repeatability scatter for our 74 clusters.  We find that the median single-epoch photometry has bright end repeatability scatter of 7.6, 7.6, 7.7 and 8.3~mmag in bands $griz$, respectively.  This compares favourably with the PS1 repeatability scatter of 16 to 19~mmag \citep{liu15} and is better than the characteristic BCS scatter of 18 to 25~mmag \citep{desai12}.  Coadd tiles with repeatability scatter larger than 20~mmag are re-examined and recalibrated to improve the photometry.

Given the large pointing offset strategy of the data acquisition for DES, each point on the sky is imaged from multiple independent portions of the focal plane.  Thus, we expect the systematic floor in the coadd photometry to scale approximately as this single epoch systematic floor divided by the square root of the number of layers contributing to the coadd.  The goal in acquiring the SV data was for it to have full DES depth, corresponding to 10 layers of imaging.  In practice the median number of exposures per band in the SV region is 6.5 to 7.5.   Thus, in principle we should expect to achieve a systematic error floor in the relative coadd photometry of around 3~mmag.  

For the analyses presented below we use {\tt mag\_auto} as our estimate of the total galaxy magnitude, and we use {\tt mag\_detmodel} for galaxy colors.  The color estimator {\tt mag\_detmodel} provides an enhanced signal to noise estimate of the galaxy color that is weighted over the same, centrally concentrated and PSF corrected region of the galaxy in each band.  The shape related systematic errors in measuring galaxy colors and photometry are large compared to the systematics floor in the stellar photometry discussed above.

\begin{figure} 
\vskip-0.75cm
 \includegraphics [width=0.48\textwidth]{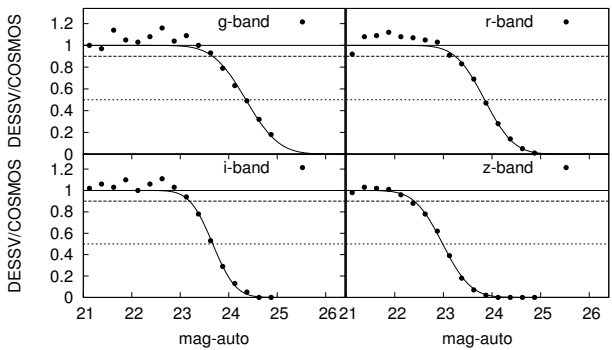}
 \vskip-0.2cm
 \caption [Completeness function for DES bands]{Completeness functions in each band for SPT-CL J0423-6143. We compare the object counts observed toward the cluster to the counts from deeper data from the COSMOS survey. The resulting completeness curve is fit by an error function, which we use to estimate 50\% and 90\% completeness.
 \label{fig:complete}}
   \vskip-0.5cm
\end{figure}


\subsection{Star-Galaxy Separation}
Our photometric catalogs are produced using model fitting photometry on homogenized coadded images, and they therefore contain two different star-galaxy separators: {\tt class\_star} and {\tt spread\_model}.  To examine the reliability of the separation we look at the values of those two classifiers as a function of magnitude \citep{desai12}.  {\tt class\_star} contains values between 0 and 1 representing a continuum between resolved and unresolved objects. At magnitudes of $\sim 20$ in the DES data the galaxy and stellar populations begin to merge, making classification with {\tt class\_star} quite noisy.  {\tt spread\_model} values exhibit a strong stellar sequence at {\tt spread\_model}$\sim 0$, whereas galaxies have more positive values. In the case of {\tt spread\_model} the two sequences start merging at roughly $\sim 22$ magnitude in each band, indicating that {\tt spread\_model} is effective at classifying objects that are approximately an order of magnitude fainter than those that are well classified by {\tt class\_star}. For this reason, we use a {\tt spread\_model} cut for the star/galaxy classification in the z-band, as it is used as a detection band. Examining the catalogs and noting the location and width of the {\tt spread\_model} stellar sequence, we find that a reliable cut to exclude stars is {\tt spread\_model}$>0.002$. 



\subsection{Completeness Estimates}
\label{sec:completeness}
Following \citet{zenteno11} we estimate the completeness of each DES-SV tile by comparing their $griz$ count histograms for all objects against those from the Cosmic Evolution Survey \citep[COSMOS; see e.g.,][]{taniguchi05}. 
COSMOS surveyed a 2~deg$^2$ equatorial field ($\alpha=150.1167, \delta=2.2058$) with the Advanced Camera for Surveys from Hubble Space Telescope (HST). These data have been supplemented by additional ground based images from the Subaru telescope, and these are the data we use here.  We extract a COSMOS source count histogram from the public photometric catalog including SDSS $griz$ bands that are transformed to our DES catalog magnitudes and normalised to the appropriate survey area.  The typical shifts in magnitude going from SDSS to DES are small compared to the size of the bins we employ to measure the count histogram.  The COSMOS counts extend down to a 10\,$\sigma$ magnitude limit of around $g\sim$25.1, $r\sim$24.9, $i\sim$25, $z\sim$24.1.    

Because the analysed DES cluster tiles do not overlap with the COSMOS footprint,  we measure the magnitude limits at 50\% and 90\% completeness levels by calculating the ratios between the DES and COSMOS area-renormalised number counts. First, we fit the DES number counts at intermediate magnitudes where both surveys are complete with a power law, whose slope is fixed to that obtained for COSMOS number counts within a similar magnitude range. Note here that we ensure that the fit is done over a magnitude range where completeness is $\ge$95\%. The ratio between these two power laws is used for renormalizing the DES number counts, effectively accounting for field-to-field variance in the counts.

We then fit an error function to the ratio of the renormalized DES and COSMOS counts to estimate the 50\% and 90\% completeness depths (see Figure~\ref{fig:complete}).  For the most part, this approach of estimating completeness works well, but due to mismatch between the power law behaviour of the counts in the region where both surveys are complete and also due to noise in the counts, the completeness estimate for DES can in some cases scatter above 1.  In particular, we have encountered some difficulties with greater stellar contamination in the DES  regions that are closest to the Large Magellanic Cloud. Thus we exclude four clusters with declination $\delta<-63^\circ$ from the fits analysis and mark them with different point styles in the figures. While the star-galaxy separation is effective at removing single stars, it fails for many binary stars that are present.

The analysis is performed using {\tt mag\_auto} and a magnitude error cut of 0.3. We use a magnitude error cut to exclude unreliable objects at the detection limit from our analysis, and we do the same thing in our analysis of the science frames.  Cutting at even larger magnitude errors does not change the depths significantly. Furthermore we exclude the cluster area within a projected distance of \Rtwohundred\  from this analysis, because this region is particularly contaminated by the presence of cluster galaxies. The mean 50\% completeness magnitude limit for the DES photometry among all 74 confirmed clusters is 24.2, 23.9, 23.3, 22.8 for $griz$, respectively, and the RMS (root mean square) variation around the mean is 0.05, 0.05, 0.05, 0.04.  The variation in completeness depths is a reminder that not all SPT cluster fields have been observed to full depth.


\begin{center}
\begin{table*}
   \caption{Properties of the SPT clusters: the cluster name, the mass $M_{200}$, the redshift $z$ and then the concentration $c_\mathrm{g}$ and number of galaxies $N_{200}$ within $R_{200}$ for the full population, the RS population and the non-RS blue population. Results are not listed where the radial profile fit does not converge.}\label{tab:properties}
    \begin{tabular}{lrcrrrrrr}
    \hline \hline   
 & \multicolumn{1}{c}{$M_{200}$}\\
 Cluster & \multicolumn{1}{c}{$[10^{14}M_\odot]$} &  $z$ &  \multicolumn{1}{c}{$c_{\mathrm{g}}$} &  \multicolumn{1}{c}{$N_{200}$} &  \multicolumn{1}{c}{$c_{\mathrm{g,RS}}$} &  \multicolumn{1}{c}{$N_{200,\mathrm{RS}}$} &  \multicolumn{1}{c}{$c_{\mathrm{g,nRS}}$} &  \multicolumn{1}{c}{$N_{\mathrm{200,nRS}}$} \\ 
\hline \hline                             
SPT-CL J0001-5440 & $  6.3 ~~^{+ 1.5 }_{-1.9}$  & $  0.89 \pm 0.03 $  &$-$& $-$ &$-$& $-$ &$-$& $-$ \\[3pt]
SPT-CL J0008-5318 &$  5.2 ~~^{+ 1.3 }_{-1.8}$  & $  0.39 \pm 0.02 $ &$-$& $-$ &$-$& $-$ &$-$& $-$ \\[3pt]
SPT-CL J0012-5352 & $  5.9 ~~^{+ 1.4}_{-2.0}$  & $  0.40 \pm 0.02 $ &$5.77^{+2.18}_{-1.46}$& $82.4^{+11.5}_{-11.5}$ &$7.75^{+2.90}_{-1.98}$& $66.7^{+9.19}_{-7.21}$ &$2.88^{+4.11}_{-1.73}$& $13.2^{+4.17}_{-8.11}$ \\[3pt]
SPT-CL J0036-4411 & $  6.1 ~~^{+ 1.4 }_{-1.8}$  & $  0.87 \pm 0.01 $  &$5.74^{+2.59}_{-2.06}$& $85.2^{+13.5}_{-12.1}$ &$10.73^{+5.65}_{-3.36}$& $48.2^{+7.36}_{-7.06}$ &$-$& $-$ \\[3pt]
SPT-CL J0040-4407 & $  17.5 ~~^{+ 2.9 }_{-3.9}$  & $  0.39 \pm 0.01 $ &$3.34^{+1.03}_{-0.89}$& $233.3^{+22.3}_{-18.4}$ &$5.45^{+1.55}_{-1.30}$& $169.8^{+14.1}_{-14.8}$ &$-$& $-$ \\[3pt]
SPT-CL J0041-4428 & $  10.2 ~~^{+ 1.7 }_{-2.4}$  & $  0.36 \pm 0.02 $&$1.15^{+0.73}_{-0.52}$& $143.5^{+9.71}_{-16.4}$ &$1.94^{+0.69}_{-0.51}$& $123.4^{+9.65}_{-11.2}$ &$-$& $-$ \\[3pt]
SPT-CL J0102-4915 & $  25.7 ~~^{+ 4.7 }_{-5.8}$  & $  0.88 \pm 0.03 $ &$2.37^{+0.88}_{-0.73}$& $223.8^{+21.5}_{-19.5}$ &$6.01^{+2.07}_{-1.48}$& $108.7^{+10.2}_{-11.3}$ &$-$& $-$ \\[3pt]
SPT-CL J0107-4855 & $  5.3 ~~^{+ 1.2 }_{-1.9}$  & $  0.60 \pm 0.02 $  &$13.97^{+10.6}_{-5.12}$& $36.2^{+9.79}_{-10.9}$ &$10.76^{+5.42}_{-3.30}$& $52.0^{+9.04}_{-8.18}$ &$-$& $-$ \\[3pt]
SPT-CL J0330-5228 & $  11.7 ~~^{+ 1.9 }_{-2.7}$  & $  0.45 \pm 0.02 $ &$-$& $-$ &$-$& $-$ &$-$& $-$ \\[3pt]
SPT-CL J0412-5106 & $  6.1 ~~^{+ 1.4 }_{-1.9}$  & $  0.28 \pm 0.03 $ &$1.25^{+0.72}_{-0.55}$& $73.3^{+11.2}_{-10.6}$ &$6.35^{+3.92}_{-2.44}$& $38.2^{+6.60}_{-6.40}$ &$-$& $-$ \\[3pt]
SPT-CL J0417-4748 & $  13.2 ~~^{+ 2.2 }_{-2.9}$  & $  0.60 \pm 0.01 $ &$0.46^{+0.33}_{-0.20}$& $149.0^{+13.1}_{-11.2}$ &$1.90^{+0.86}_{-0.55}$& $131.3^{+10.1}_{-11.8}$ &$-$& $-$ \\[3pt]
SPT-CL J0422-4608 & $  5.4 ~~^{+ 1.3 }_{-1.8}$  & $  0.67 \pm 0.02 $ &$-$& $-$ &$4.99^{+2.36}_{-1.60}$& $48.7^{+7.83}_{-8.14}$ &$-$& $-$ \\[3pt]
SPT-CL J0422-5140 & $  6.5 ~~^{+ 1.4 }_{-1.9}$  & $  0.60 \pm 0.03 $ &$-$& $-$ &$1.49^{+1.00}_{-0.48}$& $47.0^{+6.27}_{-7.42}$ &$-$& $-$ \\[3pt]
SPT-CL J0423-6143 & $  5.2 ~~^{+ 1.1 }_{-1.7}$  & $  0.63 \pm 0.02 $ &$10.99^{+6.91}_{-4.67}$& $43.8^{+10.9}_{-11.4}$ &$10.59^{+7.75}_{-4.30}$& $31.0^{+7.54}_{-6.27}$ &$-$& $-$ \\[3pt]
SPT-CL J0426-5416 & $  4.5 ~~^{+ 1.0 }_{-1.6}$  & $  1.05 \pm 0.04 $&$-$& $-$ &$1.90^{+0.40}_{-1.42}$& $20.0^{+6.45}_{-13.5}$ &$-$& $-$ \\[3pt]
SPT-CL J0426-5455 & $  8.8 ~~^{+ 1.5 }_{-2.1}$  & $  0.66 \pm 0.03 $ &$2.93^{+0.85}_{-0.58}$& $164.1^{+14.7}_{-18.1}$ &$6.31^{+3.03}_{-1.55}$& $72.1^{+9.05}_{-9.17}$ &$1.88^{+0.83}_{-0.52}$& $85.9^{+11.8}_{-14.2}$ \\[3pt]
SPT-CL J0428-6049 & $  5.5 ~~^{+ 1.4 }_{-1.8}$  & $  0.75 \pm 0.02 $  &$-$& $-$ &$-$& $-$ &$-$& $-$ \\[3pt]
SPT-CL J0429-5233 & $  5.2 ~~^{+ 1.2 }_{-1.8}$  & $  0.53 \pm 0.02 $  &$1.24^{+0.87}_{-0.44}$& $61.9^{+9.21}_{-10.5}$ &$2.65^{+1.77}_{-0.96}$& $48.1^{+7.46}_{-7.33}$ &$-$& $-$ \\[3pt]
SPT-CL J0430-6251 & $  6.3 ~~^{+ 1.5 }_{-2.0}$  & $  0.23 \pm 0.01 $  &$-$& $-$ &$1.92^{+2.01}_{-0.88}$& $19.1^{+4.45}_{-4.98}$ &$-$& $-$ \\[3pt]
SPT-CL J0431-6126 & $  7.6 ~~^{+ 1.5 }_{-2.1}$  & $  0.07 \pm 0.01 $ &$2.38^{+0.73}_{-0.53}$& $157.2^{+11.1}_{-13.7}$ &$3.95^{+1.10}_{-0.69}$& $141.7^{+9.92}_{-11.2}$ &$-$& $-$ \\[3pt]
SPT-CL J0432-6150 & $  4.3 ~~^{+ 1.0 }_{-1.6}$  & $  1.12 \pm 0.04 $ &$-$& $-$ &$-$& $-$ &$-$& $-$ \\[3pt]
SPT-CL J0433-5630 & $  5.8 ~~^{+ 1.4 }_{-1.8}$  & $  0.70 \pm 0.03 $  &$2.52^{+1.29}_{-0.85}$& $88.5^{+11.7}_{-12.9}$ &$4.93^{+3.12}_{-1.72}$& $39.4^{+6.06}_{-7.05}$ &$1.41^{+1.56}_{-0.85}$& $48.2^{+7.34}_{-13.2}$ \\[3pt]
SPT-CL J0437-5307 & $  5.4 ~~^{+ 1.2 }_{-1.8}$  & $  0.28 \pm 0.03 $  &$8.22^{+6.77}_{-4.36}$& $57.8^{+13.3}_{-12.6}$ &$9.70^{+7.82}_{-4.28}$& $39.2^{+7.79}_{-6.51}$ &$-$& $-$ \\[3pt]
SPT-CL J0438-5419 & $  18.7 ~~^{+ 3.1 }_{-4.2}$  & $  0.42 \pm 0.02 $ &$3.34^{+1.04}_{-0.83}$& $202.3^{+20.0}_{-19.5}$ &$5.78^{+1.88}_{-1.26}$& $129.6^{+12.1}_{-13.5}$ &$1.62^{+0.92}_{-0.79}$& $74.3^{+15.4}_{-12.1}$ \\[3pt]
SPT-CL J0439-4600 & $  9.3 ~~^{+ 1.6 }_{-2.2}$  & $  0.39 \pm 0.01 $  &$3.25^{+1.97}_{-1.32}$& $110.5^{+16.7}_{-14.3}$ &$5.56^{+2.53}_{-1.54}$& $81.7^{+11.8}_{-9.08}$ &$-$& $-$ \\[3pt]
SPT-CL J0439-5330 & $  6.5 ~~^{+ 1.5 }_{-1.9}$  & $  0.43 \pm 0.02 $&$6.37^{+4.99}_{-3.23}$& $64.7^{+14.5}_{-11.8}$ &$13.90^{+9.52}_{-4.18}$& $43.5^{+9.38}_{-6.80}$ &$-$& $-$ \\[3pt]
SPT-CL J0440-4657 & $  8.2 ~~^{+ 1.5 }_{-2.1}$  & $  0.40 \pm 0.01 $ &$1.53^{+1.19}_{-0.76}$& $85.6^{+11.5}_{-14.2}$ &$1.95^{+0.94}_{-0.73}$& $74.0^{+8.98}_{-9.21}$ &$-$& $-$ \\[3pt]
SPT-CL J0441-4855 & $  8.8 ~~^{+ 1.5 }_{-2.1}$  & $  0.80 \pm 0.02 $ &$-$& $-$ &$5.93^{+3.46}_{-2.17}$& $60.3^{+7.96}_{-8.14}$ &$-$& $-$ \\[3pt]
SPT-CL J0442-6138 & $  4.6 ~~^{+ 1.1 }_{-1.7}$  & $  0.95 \pm 0.04 $&$-$& $-$ &$-$& $-$ &$-$& $-$ \\[3pt]
SPT-CL J0444-4352 & $  5.7 ~~^{+ 1.4 }_{-1.9}$  & $  0.58 \pm 0.02 $&$9.51^{+5.84}_{-3.93}$& $58.7^{+12.6}_{-13.5}$ &$6.40^{+3.49}_{-2.10}$& $53.9^{+7.68}_{-8.65}$ &$-$& $-$ \\[3pt]
SPT-CL J0444-5603 & $  5.2 ~~^{+ 1.3 }_{-1.7}$  & $  0.99 \pm 0.04 $ &$7.46^{+5.75}_{-3.16}$& $53.6^{+9.19}_{-12.5}$ &$8.26^{+6.43}_{-3.26}$& $30.8^{+4.72}_{-6.27}$ &$-$& $-$ \\[3pt]
SPT-CL J0446-5849 & $  7.2 ~~^{+ 1.4 }_{-1.8}$  & $  1.11 \pm 0.03 $  &$-$& $-$ &$-$& $-$ &$-$& $-$ \\[3pt]
SPT-CL J0447-5055 & $  6.9 ~~^{+ 1.5 }_{-1.9}$  & $  0.42 \pm 0.01 $ &$10.44^{+6.52}_{-4.14}$& $76.0^{+15.6}_{-13.2}$ &$14.26^{+7.13}_{-5.06}$& $57.2^{+8.09}_{-8.93}$ &$-$& $-$ \\[3pt]
SPT-CL J0449-4901 & $  9.1 ~~^{+ 1.6 }_{-2.1}$  & $  0.80 \pm 0.02 $&$-$& $-$ &$5.37^{+2.36}_{-1.61}$& $68.4^{+8.80}_{-8.24}$ &$-$& $-$ \\[3pt]
SPT-CL J0451-4952 & $  5.6 ~~^{+ 1.4 }_{-1.9}$  & $  0.41 \pm 0.04 $  &$-$& $-$ &$4.03^{+2.01}_{-1.38}$& $62.9^{+8.47}_{-8.36}$ &$-$& $-$ \\[3pt]
SPT-CL J0452-4806 & $  5.2 ~~^{+ 1.1 }_{-1.8}$  & $  0.44 \pm 0.02 $ &$-$& $-$ &$4.92^{+4.14}_{-2.08}$& $48.7^{+6.83}_{-8.50}$ &$-$& $-$ \\[3pt]
SPT-CL J0456-4906 & $  6.3 ~~^{+ 1.4 }_{-1.8}$  & $  0.88 \pm 0.02 $&$3.68^{+2.67}_{-1.73}$& $63.6^{+11.2}_{-13.6}$ &$7.19^{+4.00}_{-2.45}$& $43.8^{+6.91}_{-7.01}$ &$-$& $-$ \\[3pt]
SPT-CL J0456-5623 & $  5.1 ~~^{+ 1.1 }_{-1.7}$  & $  0.65 \pm 0.02 $&$4.72^{+3.65}_{-2.05}$& $52.5^{+10.8}_{-12.3}$ &$3.71^{+2.77}_{-1.53}$& $40.3^{+6.29}_{-7.33}$ &$-$& $-$ \\[3pt]
SPT-CL J0456-6141 & $  5.5 ~~^{+ 1.3 }_{-1.9}$  & $  0.43 \pm 0.02 $&$10.30^{+6.38}_{-4.12}$& $66.4^{+13.7}_{-11.5}$ &$13.46^{+6.63}_{-5.00}$& $53.0^{+8.60}_{-8.09}$ &$-$& $-$ \\[3pt]
SPT-CL J0458-5741 & $  4.8 ~~^{+ 1.1 }_{-1.9}$  & $  0.20 \pm 0.01 $ &$6.83^{+4.22}_{-3.16}$& $61.6^{+11.8}_{-10.6}$ &$7.97^{+4.11}_{-3.07}$& $52.1^{+7.38}_{-7.47}$ &$-$& $-$ \\[3pt]
SPT-CL J0500-4551 & $  6.0 ~~^{+ 1.3 }_{-2.0}$  & $  0.24 \pm 0.02 $&$4.05^{+2.02}_{-1.19}$& $58.9^{+9.50}_{-9.72}$ &$3.47^{+2.05}_{-1.01}$& $26.7^{+5.13}_{-5.04}$ &$3.83^{+3.75}_{-1.65}$& $33.8^{+7.02}_{-8.73}$ \\[3pt]
SPT-CL J0500-5116 & $  7.2 ~~^{+ 1.5 }_{-2.0}$  & $  0.15 \pm 0.02 $ &$16.94^{+7.11}_{-5.41}$& $73.3^{+9.96}_{-10.7}$ &$-$& $-$ &$8.05^{+7.90}_{-3.89}$& $18.3^{+5.57}_{-6.42}$  \\ [3pt]
SPT-CL J0502-6048 & $  5.2 ~~^{+ 1.1 }_{-1.7}$  & $  0.83 \pm 0.02 $  &$0.82^{+0.89}_{-0.46}$& $43.2^{+6.74}_{-11.0}$ &$1.42^{+1.12}_{-0.58}$& $28.2^{+4.98}_{-5.49}$ &$-$& $-$ \\[3pt]
SPT-CL J0502-6113 & $  5.4 ~~^{+ 1.3 }_{-1.8}$  & $  0.80 \pm 0.02 $ &$-$& $-$ &$-$& $-$ &$-$& $-$ \\[3pt]
SPT-CL J0504-4929 & $  6.6 ~~^{+ 1.5 }_{-2.1}$  & $  0.22 \pm 0.01 $ &$4.02^{+1.67}_{-1.07}$& $84.0^{+10.2}_{-10.6}$ &$5.61^{+2.52}_{-1.56}$& $53.7^{+6.94}_{-6.97}$ &$-$& $-$ \\[3pt]
SPT-CL J0505-6145 & $  8.5 ~~^{+ 1.6 }_{-2.2}$  & $  0.29 \pm 0.01 $ &$6.40^{+3.35}_{-2.12}$& $105.7^{+15.9}_{-15.5}$ &$6.09^{+2.55}_{-1.53}$& $83.9^{+9.31}_{-8.62}$ &$-$& $-$ \\[3pt]
SPT-CL J0508-6149 & $  5.8 ~~^{+ 1.4 }_{-2.0}$  & $  0.43 \pm 0.02 $&$5.26^{+2.78}_{-1.65}$& $68.9^{+12.5}_{-12.8}$ &$3.95^{+1.68}_{-1.12}$& $52.2^{+7.63}_{-7.37}$ &$-$& $-$ \\
\hline \hline
    \end{tabular}
    \end{table*}  
\end{center}

\begin{center}
\begin{table*}
   \contcaption{Properties of the SPT clusters.}
    \begin{tabular}{lrcrrrrrr}
    \hline \hline   
 & \multicolumn{1}{c}{$M_{200}$}\\
 Cluster & \multicolumn{1}{c}{$[10^{14}M_\odot]$} &  $z$ &  \multicolumn{1}{c}{$c_{\mathrm{g}}$} &  \multicolumn{1}{c}{$N_{200}$} &  \multicolumn{1}{c}{$c_{\mathrm{g,RS}}$} &  \multicolumn{1}{c}{$N_{200,\mathrm{RS}}$} &  \multicolumn{1}{c}{$c_{\mathrm{g,nRS}}$} &  \multicolumn{1}{c}{$N_{\mathrm{200,nRS}}$} \\ 
\hline\hline                 
SPT-CL J0509-5342 & $  9.1 ~~^{+ 1.5 }_{-2.2}$  & $  0.46 \pm 0.02 $ &$-$& $-$ &$3.71^{+1.77}_{-1.18}$& $63.3^{+7.81}_{-7.95}$ &$-$& $-$ \\[3pt]
SPT-CL J0509-6118 & $  11.4 ~~^{+ 1.9 }_{-2.6}$  & $  0.40 \pm 0.03 $&$5.83^{+2.58}_{-1.94}$& $94.4^{+16.0}_{-13.1}$ &$4.55^{+1.87}_{-1.02}$& $96.2^{+9.31}_{-11.7}$ &$-$& $-$ \\[3pt]
SPT-CL J0516-5430 & $  12.3 ~~^{+ 2.0 }_{-2.8}$  & $  0.29 \pm 0.02 $ &$2.09^{+0.36}_{-0.30}$& $223.9^{+15.5}_{-13.8}$ &$3.59^{+0.75}_{-0.49}$& $154.3^{+10.4}_{-10.8}$ &$-$& $-$ \\[3pt]
SPT-CL J0516-5755 & $  5.8 ~~^{+ 1.3 }_{-1.7}$  & $  0.91 \pm 0.02 $ &$5.73^{+2.56}_{-1.70}$& $82.6^{+12.2}_{-12.9}$ &$10.64^{+5.75}_{-3.18}$& $50.1^{+6.92}_{-7.24}$ &$1.36^{+3.02}_{-1.11}$& $11.5^{+3.86}_{-8.85}$ \\[3pt]
SPT-CL J0516-6312 & $  5.9 ~~^{+ 1.4 }_{-2.0}$  & $  0.18 \pm 0.01 $ &$3.33^{+1.58}_{-1.03}$& $84.2^{+14.0}_{-14.4}$ &$2.27^{+2.34}_{-1.00}$& $19.6^{+4.69}_{-5.17}$ &$6.37^{+3.94}_{-2.65}$& $70.8^{+15.5}_{-15.4}$ \\[3pt]
SPT-CL J0517-6119 & $  7.9 ~~^{+ 1.5 }_{-2.0}$  & $  0.81 \pm 0.02 $ &$2.62^{+1.51}_{-0.92}$& $89.7^{+12.4}_{-15.8}$ &$1.92^{+1.19}_{-0.71}$& $47.9^{+6.25}_{-7.18}$ &$2.89^{+3.36}_{-1.60}$& $36.0^{+10.0}_{-13.0}$ \\[3pt]
SPT-CL J0517-6311 & $  6.3 ~~^{+ 1.5 }_{-2.0}$  & $  0.33 \pm 0.01 $ &$-$& $-$ &$2.58^{+2.54}_{-1.43}$& $17.6^{+6.34}_{-4.66}$ &$-$& $-$ \\[3pt]
SPT-CL J0529-6051 & $  6.2 ~~^{+ 1.4 }_{-1.9}$  & $  0.75 \pm 0.07 $ &$6.68^{+6.69}_{-3.75}$& $65.8^{+15.9}_{-14.3}$ &$11.7^{+9.75}_{-5.38}$& $37.1^{+8.88}_{-7.41}$ &$-$& $-$ \\[3pt]
SPT-CL J0534-5937 & $  5.2 ~~^{+ 1.1 }_{-1.7}$  & $  0.58 \pm 0.01 $ &$6.54^{+7.18}_{-3.73}$& $37.3^{+10.5}_{-11.2}$ &$5.63^{+2.85}_{-2.60}$& $31.4^{+6.49}_{-6.81}$ &$-$& $-$ \\[3pt]
SPT-CL J0539-6013 & $  5.1 ~~^{+ 1.2 }_{-1.7}$  & $  0.85 \pm 0.04 $ 	&$-$& $-$ &$-$& $-$ &$-$& $-$ \\[3pt]
SPT-CL J0540-5744 & $  7.1 ~~^{+ 1.4 }_{-1.8}$  & $  0.75 \pm 0.02 $ &$1.51^{+1.08}_{-0.61}$& $98.0^{+11.7}_{-12.9}$ &$6.90^{+4.03}_{-2.50}$& $50.3^{+9.22}_{-7.48}$ &$-$& $-$ \\[3pt]
SPT-CL J0543-6219 & $  9.5 ~~^{+ 1.6 }_{-2.3}$  & $  0.48 \pm 0.01 $ &$1.74^{+2.21}_{-1.26}$& $91.5^{+20.4}_{-21.1}$ &$6.35^{+3.07}_{-2.18}$& $59.6^{+7.62}_{-8.63}$ &$-$& $-$ \\[3pt]
SPT-CL J0546-6040 & $  5.2 ~~^{+ 1.2 }_{-1.8}$  & $  0.81 \pm 0.03 $&$8.46^{+10.5}_{-5.65}$& $28.4^{+11.1}_{-9.49}$ &$5.43^{+10.3}_{-2.66}$& $16.4^{+6.00}_{-4.80}$ &$-$& $-$ \\[3pt]
SPT-CL J0549-6205 & $  21.1 ~~^{+ 3.6 }_{-4.7}$  & $  0.42 \pm 0.02 $  &$3.11^{+0.91}_{-0.79}$& $261.1^{+27.1}_{-26.5}$ &$6.20^{+1.83}_{-1.28}$& $136.9^{+12.3}_{-13.1}$ &$1.16^{+1.10}_{-0.87}$& $113.2^{+31.7}_{-22.5}$ \\[3pt]
SPT-CL J0550-6358 & $  6.2 ~~^{+ 1.5 }_{-1.9}$  & $  0.74 \pm 0.02 $ &$-$& $-$ &$6.60^{+5.94}_{-2.73}$& $29.9^{+6.69}_{-6.92}$ &$-$& $-$ \\[3pt]
SPT-CL J0555-6406 & $  13.2 ~~^{+ 2.2 }_{-3.0}$  & $  0.40 \pm 0.02 $ &$-$& $-$ &$2.51^{+0.70}_{-0.53}$& $163.0^{+12.1}_{-14.0}$ &$-$& $-$ \\[3pt]
SPT-CL J0655-5541 & $  7.0 ~~^{+ 1.6 }_{-2.1}$  & $  0.31 \pm 0.01 $&$-$& $-$ &$6.82^{+3.03}_{-2.18}$& $67.8^{+9.74}_{-8.67}$ &$-$& $-$ \\[3pt]
SPT-CL J0658-5556 & $  28.0 ~~^{+ 5.0 }_{-6.3}$  & $  0.33 \pm 0.01 $&$2.59^{+0.77}_{-0.62}$& $145.3^{+18.5}_{-17.7}$ &$2.52^{+0.78}_{-0.49}$& $120.4^{+10.9}_{-10.5}$ &$1.04^{+0.52}_{-0.50}$& $144.6^{+29.3}_{-22.2}$ \\[3pt]
SPT-CL J2248-4431 & $  28.9 ~~^{+ 5.2 }_{-6.5}$  & $  0.37 \pm 0.02 $&$12.54^{+4.50}_{-3.56}$& $102.2^{+11.8}_{-13.3}$ &$9.37^{+4.95}_{-3.11}$& $38.1^{+6.74}_{-6.15}$ &$1.76^{+1.50}_{-1.01}$& $67.6^{+17.4}_{-17.5}$ \\[3pt]
SPT-CL J2256-5414 & $  4.7 ~~^{+ 1.0 }_{-1.6}$  & $  0.75 \pm 0.04 $&$-$& $-$ &$-$& $-$ &$-$& $-$ \\[3pt]
SPT-CL J2259-5431 & $  5.8 ~~^{+ 1.4 }_{-1.8}$  & $  0.45 \pm 0.01 $&$7.11^{+4.49}_{-3.18}$& $61.1^{+11.4}_{-13.0}$ &$8.70^{+7.58}_{-3.78}$& $37.7^{+6.92}_{-7.80}$ &$2.39^{+5.16}_{-1.52}$& $17.1^{+4.03}_{-9.67}$ \\[3pt]
SPT-CL J2300-5616 & $  6.5 ~~^{+ 1.6 }_{-2.2}$  & $  0.17 \pm 0.01 $&$8.13^{+3.14}_{-2.38}$& $91.4^{+11.0}_{-12.9}$ &$10.07^{+4.42}_{-2.88}$& $68.2^{+8.94}_{-7.73}$ &$4.48^{+5.70}_{-2.23}$& $19.6^{+6.36}_{-6.98}$ \\[3pt]
SPT-CL J2301-5546 & $  5.0 ~~^{+ 1.2 }_{-1.7}$  & $  0.76 \pm 0.02 $ &$5.33^{+3.28}_{-2.15}$& $57.1^{+10.2}_{-11.4}$ &$5.95^{+4.50}_{-2.54}$& $35.7^{+6.67}_{-7.61}$ &$2.88^{+4.48}_{-1.75}$& $14.2^{+5.72}_{-8.95}$ \\[3pt]
SPT-CL J2332-5358 & $  9.3 ~~^{+ 1.6 }_{-2.2}$  & $  0.42 \pm 0.02 $  &$4.37^{+1.75}_{-1.34}$& $97.5^{+13.3}_{-14.8}$ &$5.04^{+1.74}_{-1.34}$& $78.9^{+8.54}_{-10.5}$ &$3.36^{+4.22}_{-1.82}$& $22.5^{+7.95}_{-10.5}$ \\[3pt]
SPT-CL J2342-5411 & $  7.7 ~~^{+ 1.4 }_{-1.8}$  & $  0.98 \pm 0.05 $ &$-$& $-$ &$-$& $-$ &$-$& $-$ \\[3pt]
SPT-CL J2351-5452 & $  6.8 ~~^{+ 1.4 }_{-1.9}$  & $  0.43 \pm 0.01 $ &$7.26^{+2.69}_{-1.92}$& $112.6^{+13.6}_{-12.9}$ &$9.53^{+3.64}_{-2.39}$& $87.7^{+8.50}_{-10.8}$ &$-$& $-$ \\[3pt]
SPT-CL J2354-5633 & $  6.1 ~~^{+ 1.3 }_{-1.8}$  & $  0.55 \pm 0.01 $  &$4.94^{+1.59}_{-1.38}$& $125.1^{+14.0}_{-14.9}$ &$5.11^{+2.30}_{-1.56}$& $73.9^{+9.45}_{-9.16}$ &$4.15^{+2.40}_{-1.44}$& $51.9^{+9.70}_{-10.7}$  \\ \hline
 \hline
    \end{tabular}
    \end{table*}  
\end{center}

\begin{figure}
  \includegraphics [width=0.45\textwidth]{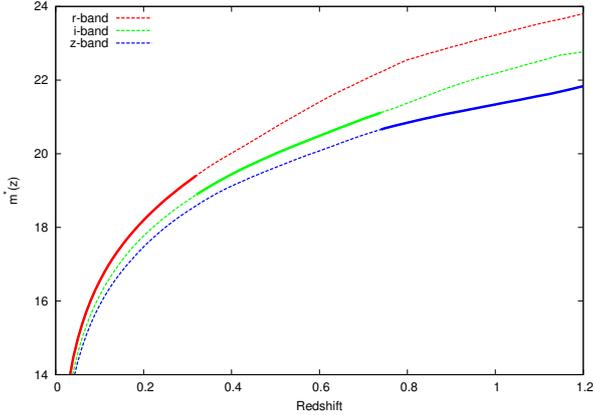}
  \vskip-0.3cm
  \caption [Characteristic magnitudes]{The characteristic magnitudes $m_*(z)$ from the CSP model that are used for this study.  In a companion paper we use measurements of $m_*$ for this cluster sample to test and calibrate this model, finding it to be an excellent description of the cluster galaxy population characteristic magnitude and its evolution with redshift. The bands are color coded and the solid lines mark the redshift ranges where each band, redward of the 4000~\AA\ break, is used.
  \label{fig:mstar}}
\end{figure}

\section{Galaxy Cluster Properties}
\label{sec:clusterproperties}
In our analysis we focus on the galaxy population within the cluster virial regions of an SPT selected sample of massive clusters.  Because the properties of the galaxy populations are a function of radius, it is important to define consistent radii for clusters of different masses and redshifts.  For the purposes of this study we adopt the region defined by \Rtwohundred-- the region where the enclosed mean density is 200 times the critical density at that redshift, and we then probe for redshift and mass trends in the ensemble within this consistently defined virial region.   To calculate \Rtwohundred\ we require a redshift and a mass estimate for each system. In this Section we describe our method for measuring the cluster redshifts and for estimating the cluster masses.


\subsection{Redshifts}
\label{sec:redshifts}
For our cluster candidates we use the RS galaxy population to estimate a photometric redshift. Our approach is similar to the one used in \citet{song12a,song12b}.  The method is based on the RS over-density in color-magnitude space.  We model the evolutionary change in color of cluster member galaxies over a large redshift range by using a composite stellar population (CSP) model. 

\subsubsection{Stellar population evolutionary model}
\label{sec:CSP}
A range of previous studies have shown that early type galaxies within clusters have stellar flux that is dominated by passively evolving stellar populations formed at redshifts $2<z<5$ \citep[e.g.,][]{bower92,ellis97,depropris99,lin06}.  We adopt a model consistent with these findings.  Specifically,  our star formation model is an exponentially decaying starburst at redshift $z = 3$ with a Chabrier IMF and a decay time of 0.4 Gyr \citep[][hereafter BC03]{bruzual03}.  We introduce tilt in the red sequence by using 6 different models, each with a different metallicity \citep{kodama97} adjusted to follow the luminosity - metallicity relation observed in Coma  \citep{poggianti01b}. Derived from the best fit metallicity-luminosity relation in \citet{poggianti01b} for Z(Hg), the corresponding metallicities used are 0.0191 $(3L^*)$, 0.0138 $(2L^*)$, 0.0107 $(L^*)$, 0.0084 $(0.5L^*)$, 0.0070 $(0.4L^*)$, 0.0047 $(0.3L^*) $.

We use DES filter transmission curves derived from the DECal system response curves that account for telescope, filters and CCDs and that include atmospheric transmission.  We use these filter transmission curves together with the \textit{EzGal} Python interface \citep{mancone12b} to create model galaxy magnitudes in the $griz$ bands and within a luminosity range of $0.3L^*<L<3L^*$. 

In the analyses that follow we measure quantities at a given redshift using the galaxy population that is brighter than $m_*+2$ where the $m_*$ is taken from this model.  In a companion paper we measure the $m_*(z)$ of the galaxy populations in this cluster sample and use those measurements to calibrate the model used here.  The $m_*(z)$ for each band is shown in Figure~\ref{fig:mstar}.  Thicker lines are used to note the redshift ranges over which the $m_*$ is used for a particular band; we shift bands with redshift in an attempt to employ a similar portion of the rest frame spectrum at all redshift.

\subsubsection{Redshift Measurements}`
\label{sec:redshift_measurements}
A cluster is confirmed by identifying an excess of RS galaxies at a particular location in color space corresponding to the redshift of the cluster.  We scan through redshift examining the galaxy population within a particular projected region.  Following previous work on X-ray and SZE selected clusters\citep{song12a,song12b}, we define a search aperture for each cluster that is centered on the SPT candidate position and has a radius of $0.5*$\Rtwohundred, which is calculated using the SZE mass proxy (see discussion in Section~\ref{sec:masses}). To measure the number of galaxies above background at each redshift, we adopt a magnitude cut of $0.4 L^{*}$ together with a magnitude uncertainty cut $\sigma_{mag}<0.1$ to exclude faint galaxies.  Each galaxy within the radial aperture is assigned two different weighting factors: one accounting for the spatial position in the cluster area and one for the galaxy position in color-magnitude space.  The color weighting $L_{\mathrm{col}}$ accounts for the orthogonal distance $d$ of each galaxy in color-magnitude space from the tilted RS appropriate for the redshift being tested and has a Gaussian form:
\begin{equation}
   L_{\mathrm{col}}= \exp{ (- \frac{d^2}{2\sigma_{\mathrm{col}}^2})}
\label{eq:RS}
\end{equation}
Here $\sigma_{\mathrm{col}}^2 = \sigma_{\mathrm{int}}^2+\sigma_{\mathrm{proj}}^2$, where we adopt $\sigma_{\mathrm{int}}=0.05$ as the intrinsic scatter in the RS (initially assumed to be fixed) and $\sigma_{\mathrm{proj}}^2$ is the combined color and magnitude measurement uncertainty projected on the orthogonal distance to the RS.   The spatial weighting
\begin{equation}
 L_{\mathrm{pos}} \sim \frac{1}{x^2-1} f(x)
 \label{eq:SW}
 \end{equation}
has the form of the projected NFW profile \citep{navarro97} and the profile is described in detail in Section~\ref{sec:profile}. The final weighting is the product of both factors.

In this way, all galaxies close to the cluster center and with colors consistent with the red sequence at the redshift being tested are given a high weight, whereas galaxies in the cluster outskirts with colors inconsistent with the red sequence are given a small weight. We use a local background annulus within $\sim 1.5-3$\Rtwohundred, depending on the extent of the tile, to define the background region for statistical background correction. The background measurement is obtained by applying for each galaxy the color weight and a mean NFW weight derived from the cluster galaxies and then correcting for the difference in area.

We observe the color magnitude relation using the photometric band that contains the rest frame 4000~\AA\ break and another band redward of this.  The appropriate colors for low redshift clusters $z<0.35$ are $g-r$ and $g-i$, for intermediate redshift clusters $0.35<z<0.75$ are $r-i$ and $r-z$ and for clusters at redshifts $z>0.75$ are $r-z$ and $i-z$.  These colors provide the best opportunity to separate red from blue galaxies as a proxy for passive and star forming galaxies, respectively, given the depth constraints of our survey data.  For each of these color combinations we construct histograms of the weighted number of galaxies as a function of redshift. The weighted number of galaxies is defined as the sum of all galaxy weights within the cluster search aperture that has been statistically background subtracted. The cluster photometric redshift is then estimated from the most significant peak in the histogram.  The photometric redshift uncertainty is the 1$\sigma$ positional uncertainty of the peak, which is derived by fitting a Gaussian to the peak, and then dividing the $\sigma$ width of the  peak by the square-root of the weighted galaxy number at the peak. 

To test our photometric redshifts we use a sample of 20 spectroscopic redshifts available in the literature \citep{song12b,ruel14}.  A fit to the differences between the photometric and spectroscopic redshifts that allows for a linear redshift dependence and offset provides a slope that is consistent with unity; fitting only an offset indicates that the photometric redshifts are typically high by $0.019\pm0.004$.  The RMS scatter of $\Delta z /(1+z)$ using our small spectroscopic cluster sample is 0.02.  Thus, the cluster photometric redshift performance is consistent with our expectation from studies of other SPT selected cluster samples \citep[][]{song12b,bleem15} .  Given the scale of the bias in our photo-z's, we do not apply corrections.  Photometric redshift biases at this level are not relevant for the analyses that follow.


The redshifts for all the confirmed clusters are listed in Table~\ref{tab:properties}. The mean redshift of our cluster sample is 0.56, the median is 0.46, and the sample lies between 0.07 and 1.12.  For redshifts $z>1$ it is better to use the optical data in combination with NIR data to estimate reliable photometric redshifts;  nevertheless, with the few clusters we have in this redshift range our DES photometric redshifts provide no evidence for large errors. 


\begin{figure} 
 \includegraphics [width=0.48\textwidth]{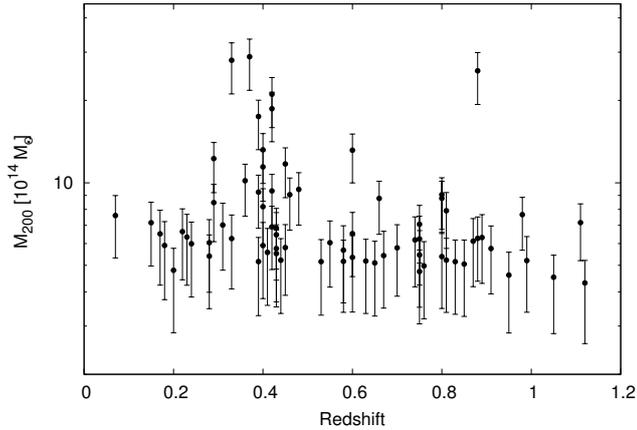}
 \vskip-0.2cm
 \caption[Mass distribution of the SPT cluster sample]{The cluster sample as a function of mass $M_{200}$.  The error bars reflect the 1$\sigma$ mass uncertainties.  The median mass of the sample is $6 \times 10^{14} M_{\odot}$ and the median redshift is $z=0.46$.  We adopt these median values as pivot points in our joint mass and redshift power law fits to the observed galaxy population properties. Note that the typical error for the photometric redshifts is $\sim0.02$.
\label{fig:mass-redshift}}  
\vskip-0.25cm
\end{figure}

\subsection{Cluster Masses}
\label{sec:masses}
The SPT-SZ survey consists of mm-wave imaging of 2500~deg$^2$ of the southern sky in three frequencies (95, 150 and 220~GHz) \citep[e.g.,][]{story13}.  Details of the survey and data processing are published in \citet{schaffer11}. Galaxy clusters are detected via their thermal SZE signature in the 95 and 150 GHz SPT maps using a multi-scale and multi-frequency matched-filter approach \citep{melin06,vanderlinde10}.  This filtering produces a list of cluster candidates, each with a position and a detection significance $\xi$, which is chosen from the filter scale that maximises the cluster significance.  We use this selection observable also as our mass proxy.

Due to observational noise and the noise biases associated with searching for peaks as a function of sky position and filter scale, we introduce a second unbiased SZE significance $\zeta$, which is related to the mass $M_{500}$ in the following manner: 
\begin{equation}
  \zeta = A_{\mathrm{SZ}}\left(\frac{M_{500}}{3\times 10^{14}M_{\odot}h^{-1}}\right)^{B_{\mathrm{SZ}}}\left(\frac{E(z)}{E(0.6)}\right)^{C_{\mathrm{SZ}}}
\label{eq:mass}
\end{equation}
where $A_{\mathrm{SZ}}$ is the normalization, $B_{\mathrm{SZ}}$ is the slope and $C_{\mathrm{SZ}}$ is the redshift evolution parameter.  An additional parameter $D_{\mathrm{SZ}}$ describes the intrinsic log-normal scatter in $\zeta$ at fixed mass, which is assumed to be constant as a function of mass and redshift. For $\xi>2$, the relationship between the observed $\xi$ and the unbiased $\zeta$ is
\begin{equation} \label{eq:xizeta}
  \zeta = \sqrt{\langle\xi\rangle^2-3}.
\end{equation}

For our analysis we use the masses from the recent SPT mass calibration and cosmological analysis \citep{bocquet15} that uses a 100 cluster sample together with 63 cluster velocity dispersions \citep{ruel14} and 16 X-ray $Y_X$ measurements \citep{andersson11,benson13}.  The \citet{bocquet15} analysis combines this SPT cluster dataset with CMB anisotropy constraints from WMAP9 \citep{hinshaw13}, distance measurements from supernovae \citep{suzuki12} and observations of baryon acoustic oscillations \citep{beutler11,padmanabhan12,anderson12}.  

In summary, the mass estimates (and associated uncertainties) for each cluster include bias corrections associated with selection (the so-called Eddington bias) and are marginalized over cosmological and scaling relation parameters.  The conversion from the $M_{500}$ in Equation~\ref{eq:mass} to the $M_{200}$ used here assumes an NFW model \citep{navarro97} with a concentration $c$ sampled from structure formation simulations \citep{duffy08}.  The cluster masses are listed in Table~\ref{tab:properties}, and the mass--redshift distribution for the full cluster sample is shown in Figure~\ref{fig:mass-redshift}.

All the details of the mass calibration can be found in \citet{bocquet15}.  For the purposes of this work we note that if we had adopted the {\it Planck} CMB anisotropy constraints instead of WMAP9 it would increase our masses by $\sim$6\%.  Also, our characteristic cluster mass uncertainty is $\sim$20\%, corresponding to a virial radius uncertainty of $\sim$7\%.

\begin{table}
\begin{center}
\caption{Definition of redshift bins. We list the bin number, the redshift range , the depth in terms of $m_*$, the number of contributing clusters as well as the bands used for the color and the magnitude.}\label{tab:stacks}
\begin{tabular}{lccccc}
    \hline \hline   
\# & z & Depth & $N_{clu}$ & Color & Band\\ \hline \hline                             
1 &0.07-0.23  & $m_*+2$ & 7   &g-r  &r\\[3pt]
2 &0.24-0.33  & $m_*+2$ & 7    &g-r &r\\[3pt]
3 &0.33-0.42  & $m_*+2$ & 12   &r-i  &i\\[3pt]
4 &0.42-0.48  & $m_*+2$ & 11   &r-i  &i\\[3pt]
5 &0.53-0.70  & $m_*+2$ & 12   &r-i &i\\[3pt]
6 &0.74-0.80  & $m_*+1.7$ & 8   &i-z &z\\[3pt]
7 &0.80-0.88  & $m_*+1.7$ & 8   &i-z &z\\[3pt]
8 &0.89-1.12  & $m_*+1.2$ &  8  &i-z &z\\  \hline
\hline
\end{tabular}
\end{center}
\end{table}


\section{Galaxy Population Properties}
\label{sec:galaxypopulations}

In Section~\ref{sec:red_pop} we study the color distributions of our cluster galaxies to test whether our fiducial CSP model is a good description of the data and to explore whether the RS population is evolving over cosmic time.  In Sections~\ref{sec:profile} and \ref{sec:n200}, we examine the radial distribution of cluster galaxies and the halo occupation number, respectively.  Section~\ref{sec:redfraction} contains a study of the red fraction and its dependence on mass and redshift.

\begin{figure} 
\vskip-0.3in
\hskip-0.1in
\includegraphics [width=0.52\textwidth]{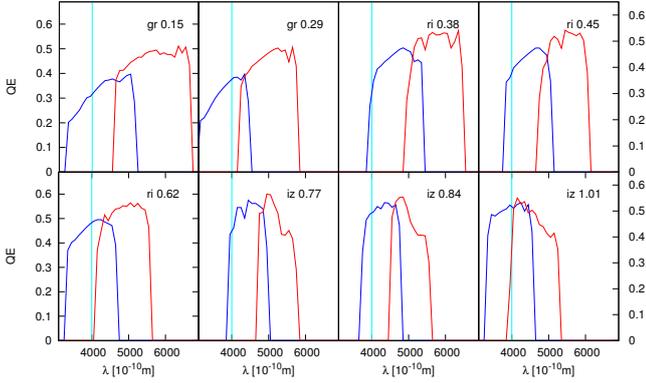}
\vskip-0.2in
\caption [Rest frame coverage]{Plots of the average relative throughput or quantum efficiency (QE) in the rest frame of the blue and red bands we use within the 8 different redshift bins defined in Table~\ref{tab:stacks}. Redshifts and observed band combinations are noted in each panel, and 4000~\AA\ is marked in cyan.  By adopting different band combinations as a function of redshift we are able to probe similar portions of the rest frame SED over the full redshift range.
\label{fig:restframe}}
\end{figure}


\subsection{Red Sequence Selection and Evolution}
\label{sec:red_pop}
We wish to be able to study both the RS and non-RS galaxy populations as a function of redshift, and doing so means that we need to have a reliable way of selecting one or the other.  In the simplest case this means we need to know the typical color, tilt and width of the RS as a function of redshift.  We have already shown in Section~\ref{sec:redshift_measurements} that our fiducial CSP model produces photometric redshifts with small biases;  this serves as a confirmation that the color evolution of the RS in our CSP model is consistent with that in our cluster sample out to $z\sim1.1$.  To test the RS tilt and measure any evolution in width we combine information from subsamples of $\sim10$ clusters each within redshift bins and use these stacks to test our model.  Stacking the clusters helps to overcome the Poisson noise in the color-magnitude distribution of any single cluster, allowing the underlying color distribution of the galaxies to be studied more precisely.

\subsubsection{Galaxy Color Distributions}
\label{sec:colordistributions}

Table~\ref{tab:stacks} contains a description of the different redshift bins within which we stack the clusters.  The table shows the redshift range of the clusters in the bin, the depth to which we are able to study the color-magnitude distribution, the number of clusters in each bin, and the color and band combinations used.  We attempt to study the color distribution to a fixed depth corresponding to two magnitudes fainter than the characteristic magnitude $m_*(z)$.  However, given the depth of the DES-SV data we are only able to study galaxies to $m_*+1.7$ at $z>0.7$ and to $m_*+1.2$ at $z>0.9$.  

We also wish to study the same portion of the spectral energy distribution (SED) in each redshift bin.  Given the broad band photometry at our disposal this is not formally possible.  Nevertheless, we attempt to minimize the impact of band shifts within the rest frame focusing in each bin on the band containing the 4000~\AA\ break and the band redward of that band.  Here again, in the highest redshift bin we have to compromise and use $i-z$ even though a more appropriate band combination would be $z-J$.   Figure~\ref{fig:restframe} contains the effective rest frame coverage of our bands as averaged over the specific clusters in that bin.  It is clear that even when shifting band combinations with redshift, the rest frame coverage varies considerably (e.g. compare bin 2 with bin 3 in Figure~\ref{fig:restframe}).  We account for this variation when interpreting the width of the red sequence in Section~\ref{sec:RSwidth}.

\begin{figure*} 
\vbox to 2.7in{\vskip-0.25in\hbox to \hsize{\hskip-0.2in
  \includegraphics [width=0.58\textwidth]{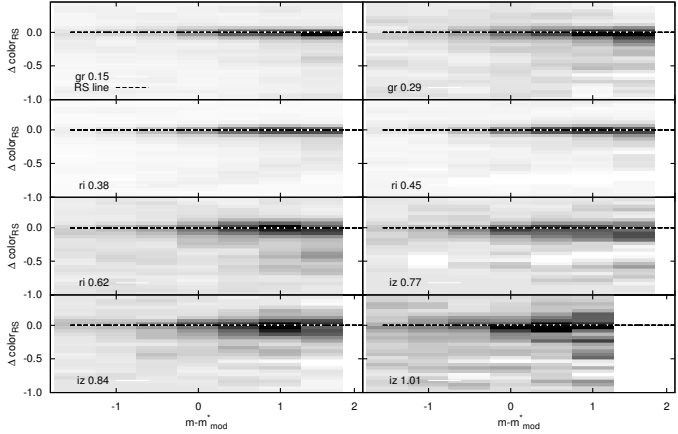}\hfil}
  \vfil}
 \vskip-3.15in
 \vbox to 2.7in{\hbox to \hsize{\hfil
  \includegraphics [width=0.47\textwidth]{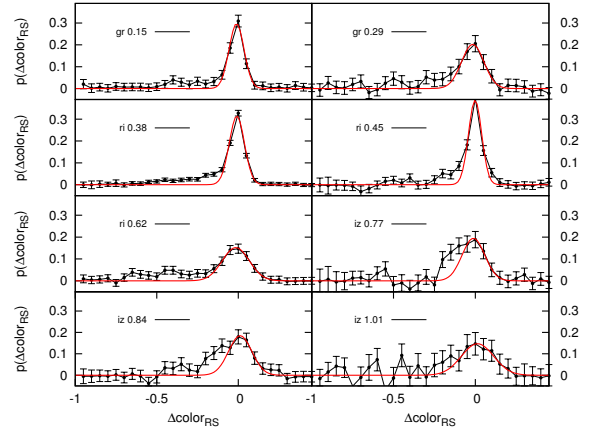}}
  \vfil}
  \vskip-0.1in
 \caption [Stacked color-magnitude and color distributions for 8 different redshift bins]{Stacked cluster galaxy color--magnitude distributions (left) for the 8 different redshift bins (see Table~\ref{tab:stacks}).  The magnitude scale is defined relative to the $m_*(z)$ of our passive evolution model, and the color offset is defined with respect to our tilted RS model (see Section~\ref{sec:CSP}). A common grey scale across all bins represents the completeness corrected and background subtracted number density of galaxies per magnitude and color bin.  The RS is clearly apparent at all redshifts, extending cleanly to $m_*+2$ in the lower redshift bins.  On the right is the stacked galaxy color distribution for the same redshift bins.  All distributions are normalized to unit area.  At higher redshift the distribution of galaxies bluer than the RS grows more prominent, and the RS has lower contrast.  
  \label{fig:RSstack} }
  \vskip-0.25cm
\end{figure*}

To construct the individual cluster color-magnitude distributions, we measure the color of each galaxy relative to the color of the tilted RS at that redshift and its magnitude relative to the characteristic magnitude $m_*(z)$.  We combine all galaxies that lie within a projected radius \Rtwohundred\ and make a statistical background correction using the local background region inside an annulus of $1.5 - 3R_{200}$. The stacked color distribution is then the average of the color distributions of the individual clusters in the bin; we normalize this distribution in each bin.

The resulting normalized and stacked color-magnitude distributions are shown in Figure~\ref{fig:RSstack} (left).   The locations of high cluster galaxy density are shown in black and low density in white.  All eight redshift bins have the same greyscale color range, allowing one to compare galaxy densities not only within a bin but also across bins.  The location of the RS as defined by our CSP model lies along the line where the color difference with the RS is zero.  In all panels there is a strong RS with an associated bluer non-RS galaxy population.  The contrast of the RS drops with redshift (note here that this drop is not due to incompleteness at that depth, as we correct each individual cluster according to its completeness as described in Section~\ref{sec:completeness}). The observed contrast is sensitive both to the cluster galaxy population and the density of background galaxies in the relevant locations of color-magnitude space.  Beyond $z\sim0.6$ there appears to be a more significant blue population than in the lower redshift bins \citep[see also][]{loh08}.  This is due to possible evolution of the red fraction, which we come back to in Section~\ref{sec:redfraction}.  In addition, the RS population extends over a range of magnitude to $m_*+2$ in the lower redshift bins but shows up less strongly at the faintest magnitudes in the higher redshift bins.  Note that over the full redshift range there is no apparent tilt of the stacked color-magnitude distribution with respect to the tilt of our CSP model.

To further increase the signal to noise ratio to study the color distribution of the cluster galaxies, we integrate these distributions over magnitude.  Figure~\ref{fig:RSstack} (right) contains these projected galaxy color distributions in each of the eight redshift bins.  Points show the relative galaxy number density and the RS is modeled as a Gaussian in red. We find that the offset of the RS Gaussian is consistent with 0 within $1\sigma$ in all of the redshift bins. Thus, the RS Gaussian has a color consistent with our CSP model (see Section~\ref{sec:CSP}). The observed width of the RS Gaussian increases to higher redshift, and its contrast relative to the non-RS galaxy population falls.  As examined in Section~\ref{sec:RSwidth}, this growth in RS width is driven by the increased color measurement uncertainty in the fainter galaxies together with some potential increase in its intrinsic width.  The RS population is dominant at lower redshift, where the non-RS galaxies appear as an "extended wing" to the RS population, and at redshifts $z\simeq0.77$ the non-RS and RS populations become less easily distinguishable.

\subsubsection{Red Sequence Selection}
\label{sec:RSselection}
In the analyses that follow we examine the RS and non-RS populations.  When examining the RS population we assign an individual galaxy $i$ in the $j$-th redshift bin a likelihood $P(c_i,z_j)$ of being a RS member that depends on its color $c_i$ and on the color distribution from the corresponding stack:
\begin{equation}
	P(c_\mathrm{i},z_\mathrm{j}) = \frac{A(z_\mathrm{j}) \exp{- \frac{\left(c_\mathrm{i}-c(z_\mathrm{j})\right)^2}{2\sigma(z_\mathrm{j})^2}} }{P_{\mathrm{obs}}(c_\mathrm{i},z_\mathrm{j})}
	 \label{eq:red}
\end{equation}
where  $A(z_{\mathrm{j}})$, $\sigma(z_{\mathrm{j}})$ and $c(z_{\mathrm{j}})$ denote the amplitude, width and color offset of the RS Gaussian in redshift bin $z_\mathrm{j}$.  $P_{\mathrm{obs}}(c_\mathrm{i},z_\mathrm{j})$ denotes the observed color distribution at the given color $c_\mathrm{i}$ and in the j-th redshift bin. In the analyses that follow each galaxy is weighted with this probability, enabling us to carry out a meaningful study of the RS population over a broad redshift range accounting for variation in intrinsic scatter and changes in the color measurement uncertainties.

\begin{figure} 
\begin{center}
\vskip-0.00cm
  \includegraphics [width=0.45\textwidth]{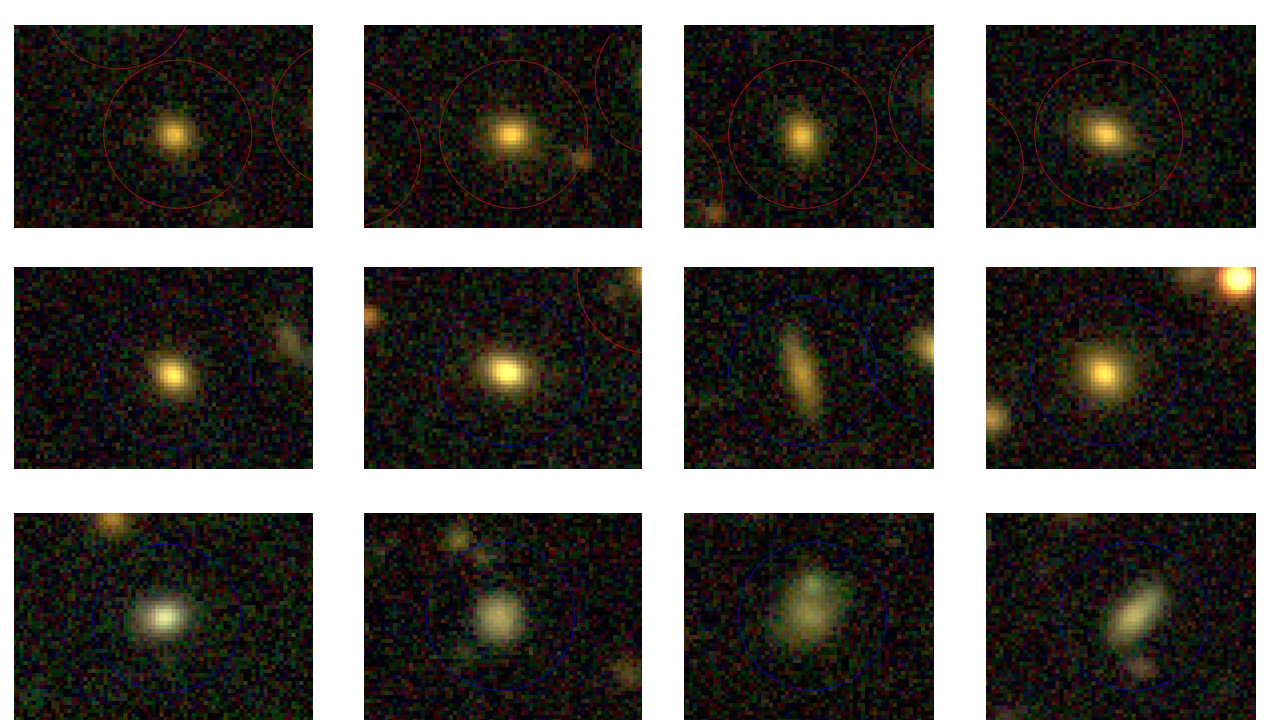} \end{center}
  \vskip-0.1cm
  \caption [Optical image gallery of RS selected galaxies]{Image gallery of galaxies at $R<R_{200}$ within the field of SPT-CL J2351-5452. The top row contains 4 examples of galaxies with a high likelihood ($\ge90\%$) of being RS members, and the second row shows galaxies with intermediate likelihood ($\sim40\%$).  We note that these constitute a population of galaxies whose colors lie between those of the RS and the bluer spirals.  The third row contains galaxies with a low likelihood ($\le20\%$) of being RS members. The majority of these are bluer disk galaxies.}
  \label{fig:image2}
  \vskip-0.2cm
\end{figure}

To examine our RS selection we create pseudo-color images of our color-selected galaxy population as shown in Figure~\ref{fig:image2}.  The top row of Figure~\ref{fig:image2} marks galaxies with likelihood $P(c_\mathrm{i},z_\mathrm{j})\ge90\%$ to be part of the cluster RS population. These galaxies all have similar red colors.  In contrast, the bottom row contains galaxies with a likelihood $P(c_\mathrm{i},z_\mathrm{j})\le20\%$ of being RS members.  These are typically blue spirals.  The middle row contains galaxies with likelihoods of $P(c_\mathrm{i},z_\mathrm{j})\sim40\%$ of being RS galaxies.  They have colors that place them between the RS and the newly infalling spirals from the field.  

The behavior seen in SPT-CL J2351-5452 (Figure~\ref{fig:image2}) is similar to that in other clusters in our sample.  Thus, through visual inspection of our sample we confirm that the color selection based on the projected color stacks in Section~ \ref{sec:red_pop} is reasonably separating the RS population from the blue cluster population. 

%
\begin{figure}
  \includegraphics [width=0.48\textwidth]{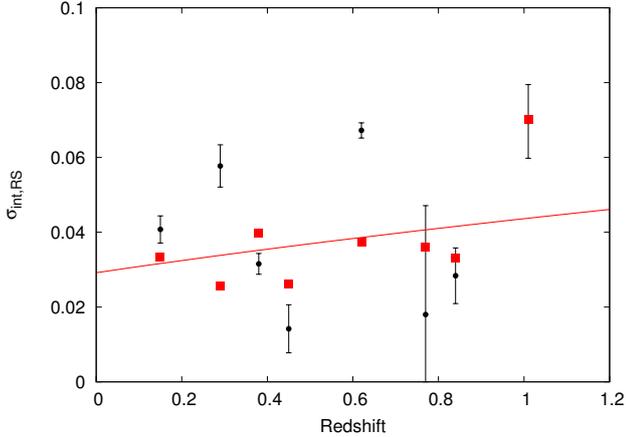}
  \vskip-0.1in
  \caption [Evolution of the intrinsic width of the Red Sequence]{Black points mark the color measurement-noise corrected widths of the RS at each redshift using the band combinations in Table~\ref{tab:stacks}.  The red points are the estimated RS widths in the rest frame $g-r$ color, given the width in the observed band combination.  There is no compelling evidence for RS width evolution within the rest frame bands (see Equation~\ref{eq:rs_intr}).
  \label{fig:evo_RS} }
\end{figure}
%


\subsubsection{Red Sequence Intrinsic Width}
\label{sec:RSwidth}

The stacked color distributions in Figure~\ref{fig:RSstack} also provide constraints on the change of the intrinsic scatter of the RS with redshift.  The RS width reflects the diversity of the stellar populations (metallicity, age and star formation history) and extinction within the passively evolving component of the cluster galaxy population.  Often the width of the red sequence is interpreted only in terms of constraints on the age variation in the stellar populations \citep[e.g.,][]{kodama97,bernardi05,gallazzi06}.   

To extract the intrinsic scatter we determine the color measurement uncertainty $\sigma_{\mathrm{col,j}}$ to produce an estimate of the intrinsic width $\sigma_{\mathrm{int,j}}$ of the RS within the j-th redshift bin 
\begin{equation}
	\sigma_{\mathrm{int,j}}^2 =\sigma_{\mathrm{j}}^2- \sigma_{\mathrm{col,j}}^2
\end{equation}
where $\sigma_{\mathrm{j}}$ is the observed RS width in redshift bin j.  To calibrate the measurement error in {\tt mag\_detmodel} colors for our coadd catalogs and to calculate  $\sigma_{\mathrm{col,j}}^2$, we examine how repeated color measurements of the same objects compare.  To do this we create two approximately full depth coadd tiles at the same sky position using different sets of single epoch exposures.  Each of these coadd tiles has about 10 exposures in each band and is therefore consistent with a full depth DES tile like those available for our sample.  Our test tiles draw upon the DES-SNe survey imaging dataset, where fields have been repeatedly observed over long periods to identify new SNe. With two independent tiles of the same sky region, we then compare the {\tt mag\_detmodel} colors from the two tiles as a function of {\tt mag\_auto} (e.g. we compare the {\tt mag\_detmodel} colors $g-r$ as a function of {\tt mag\_auto} $r$) . Fitting a simple power law relation, we model the variance in the {\tt mag\_detmodel} measurements as a function of magnitude $\sigma_{mod}(mag_i)$. 

Because the photometric depths of the coadd tiles in our cluster sample do vary somewhat, we determine the $10\sigma$ depth for the two coadd tiles as well as for all the cluster tiles. For this purpose we adopt the magnitude at which the median magnitude error in {\tt mag\_auto} equals 0.1. To account for the depth difference between the SNe field and the typical cluster field, we fit the power law as a function of $m-m_{10\sigma}$, where $m_{10\sigma}$ denotes the $10\sigma$ depth in the SNe field.  Essentially we are then measuring the {\tt mag\_detmodel} color scatter with respect to the $10\sigma$ {\tt mag\_auto} depth in the SNe field and then applying that as a model of the color scatter in each cluster field.

Because we are analyzing color stacks within redshift bins, we determine the mean $10\sigma$ depth of the clusters contributing in the bin.  From the color stacks in Figure~\ref{fig:RSstack} we have a measure of the number of galaxies in magnitude bins within the magnitude range between $m_*-2$ and $\sim m_*+2$.  We define this as $N(mag_\mathrm{i})$ where $mag_\mathrm{i}$ is the magnitude associated with bin $i$. Then we estimate the measurement contribution to the color width as:
\begin{equation}
	\sigma^2_{\mathrm{col,j}} = 
	{{\Sigma_\mathrm{i} {N(mag_\mathrm{i})*\sigma^2_{\mathrm{mod}}(mag_\mathrm{i}-m_{10\sigma,\mathrm{j}})}} 
	\over 
	{\Sigma_\mathrm{i} {N(mag_\mathrm{i})}}}
\end{equation}
where $m_{{10\sigma,\mathrm{j}}}$ denotes the mean $10\sigma$ depth of all clusters within the j-th redshift bin. Thus the final estimate of the color measurement variance is basically a weighted sum of the color measurement variances as a function of magnitude. 

The color measurement noise-corrected intrinsic widths of the RS are plotted as black points in Figure~\ref{fig:evo_RS}.  Note that we are measuring RS scatter in $g-r$, $r-i$ and $i-z$ over this redshift range, which allows us to probe similar but not identical portions of the rest frame spectrum of these galaxies.  To estimate the scatter within the rest frame $g-r$ color, we build a library of model SEDs using the code \texttt{GALAXEV} from \citet{bruzual03}.  We use exponentially declining star formation histories with different decay times \citep[see][]{chiu16a}.  For each model with a different decay time we obtain predicted colors for $\sim 200$ different stellar population ages. In total our library contains around 5000 template SEDs. Then for each redshift bin we extract template SEDs in the observed frame color that reproduce the observed Gaussian of the RS with a mean $m_*$ color and intrinsic width. For this set of individual SEDs we then extract the distribution of rest frame $g-r$ color and fit this to a Gaussian. We mark the estimated rest frame $g-r$ scatter with red squares in Figure~\ref{fig:evo_RS}.  To simplify the figure, the measurement uncertainties are placed only on the black points.



We fit a power law to the RS width as a function of redshift and find
\begin{equation}
  \sigma_{\mathrm{int,RS}}= (0.036 \pm 0.002)  \left(\frac{1+z}{1+0.46}\right)^{(0.58\pm0.47)}
  \label{eq:rs_intr}
\end{equation}
Thus, our current sample indicates a characteristic RS width of ($36\pm2$)~mmag in the rest frame $g-r$ color at the pivot redshift $z=0.46$ of our sample and provides no statistically significant evidence of an increase in width as a function of redshift  after increasing measurement uncertainties are accounted for.  

RS scatter has been studied previously \citep[e.g.,][]{aragon-salamanca93,stanford98,blakeslee03,mei06,mei09}. As these studies correct to a restframe $U-V$ color and do not report a redshift trend on their individual cluster data, we restrict ourselves to a qualitative comparison. Summarized in \citet{papovich10}, the RS scatter shows typical values of $\sim$25\,mmag at $z\sim0$ and increases towards $\sim$140\,mmag at redshift 1.62; this trend would be consistent with the observations we present above if much of the width increase occurs at redshifts $z>0.9$. 


\begin{figure} 
\vskip-0.15in
  \includegraphics [width=0.48\textwidth]{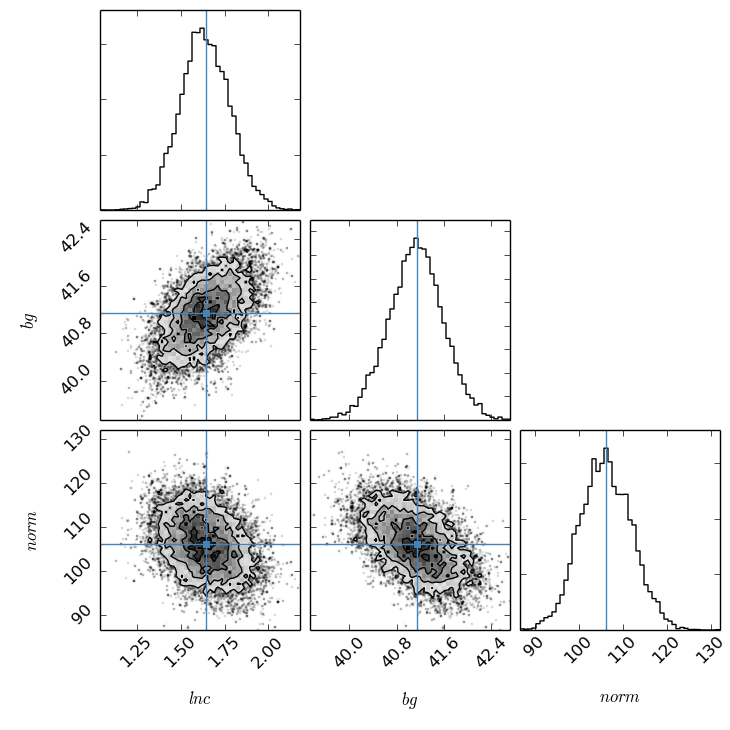}
  \vskip-0.3cm
  \caption [Stacked Radial Profile Constraints from MCMC for the full population]{Parameter constraints for the stacked cluster profile using the full population within the redshift range $0.07<z<0.23$. We show the joint  and fully marginalized constraints in $ln(c)$, the background $bg$ and the number of galaxies $norm$ projected within \Rtwohundred. The best fit is marked with the vertical blue lines. } 
  \label{fig:constrain}
  \vskip-0.2cm
\end{figure}

\begin{figure} 
\vskip-0.5cm
  \includegraphics [width=0.48\textwidth]{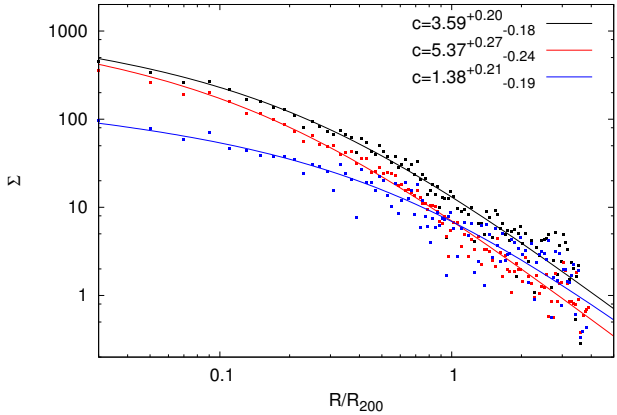}
  \vskip-0.3cm
  \caption [Stacked Radial Profile of all clusters in the sample]{The black, red and blue points (and lines) show the stacked, background subtracted radial profiles and their best fit NFW models for the full, RS and blue non-RS populations, respectively. We show the quantity $\Sigma = N/(A_{200})$, where $A_{200}$ is the projected virial area.  These stacks contain all clusters in the sample except for SPT-CL J0330-5228, which has no i-band coverage.  All individual profiles extend to $4R_{200}$.  The RS and blue non-RS populations are selected using the observed color distributions in eight redshift bins (see Equation~\ref{eq:red}).  The RS population is strongly clustered, whereas the blue non-RS population is clustered but with a lower concentration.  
  } 
  \label{fig:RP_stack}
  \vskip-0.2cm
\end{figure}

\begin{figure*} 
\vskip-1.5cm
\hskip-0.30in
  \includegraphics [width=1.05\textwidth]{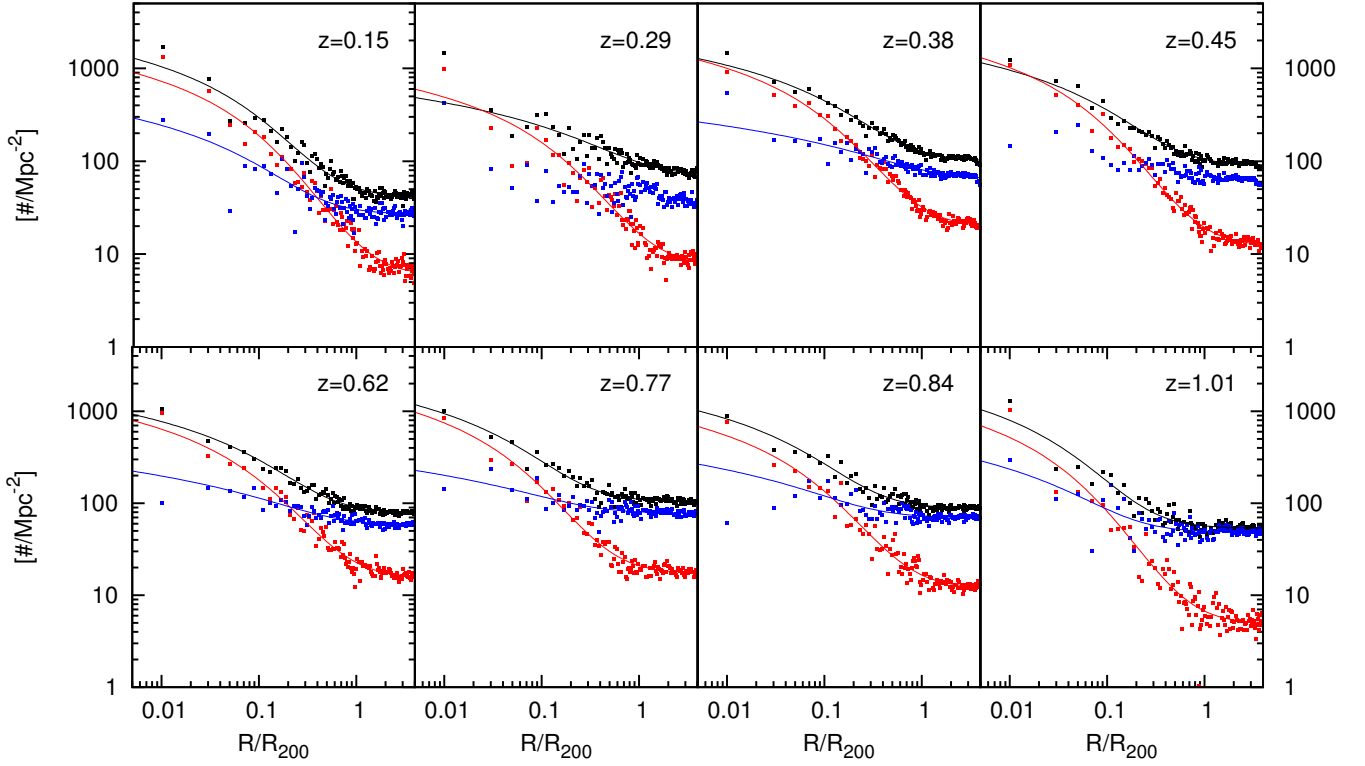}
  \vskip-1.0cm
  \caption [Stacked Radial Profiles of clusters within 8 different redshift bins]{The color coding is the same as in Figure~\ref{fig:RP_stack}, but here the stacks without background subtraction are shown for the eight different redshift bins. Again, all profiles extend to $4R_{200}$.  In each redshift bin we find that the blue non-RS population is concentrated toward the cluster center but with lower concentration than the RS population.  The concentrations from the stacks show no statistically significant mass or redshift trends (see Table~\ref{tab:fitvalues}).  No model is shown for the blue non-RS population in redshift bins 2 and 4, because the concentration is not constrained.} 
  \label{fig:RP_stack_bins}
  \vskip-0.2cm
\end{figure*}

\begin{figure*} 
\vskip-0.5cm\hskip-0.05in
  \includegraphics [width=1.00\textwidth]{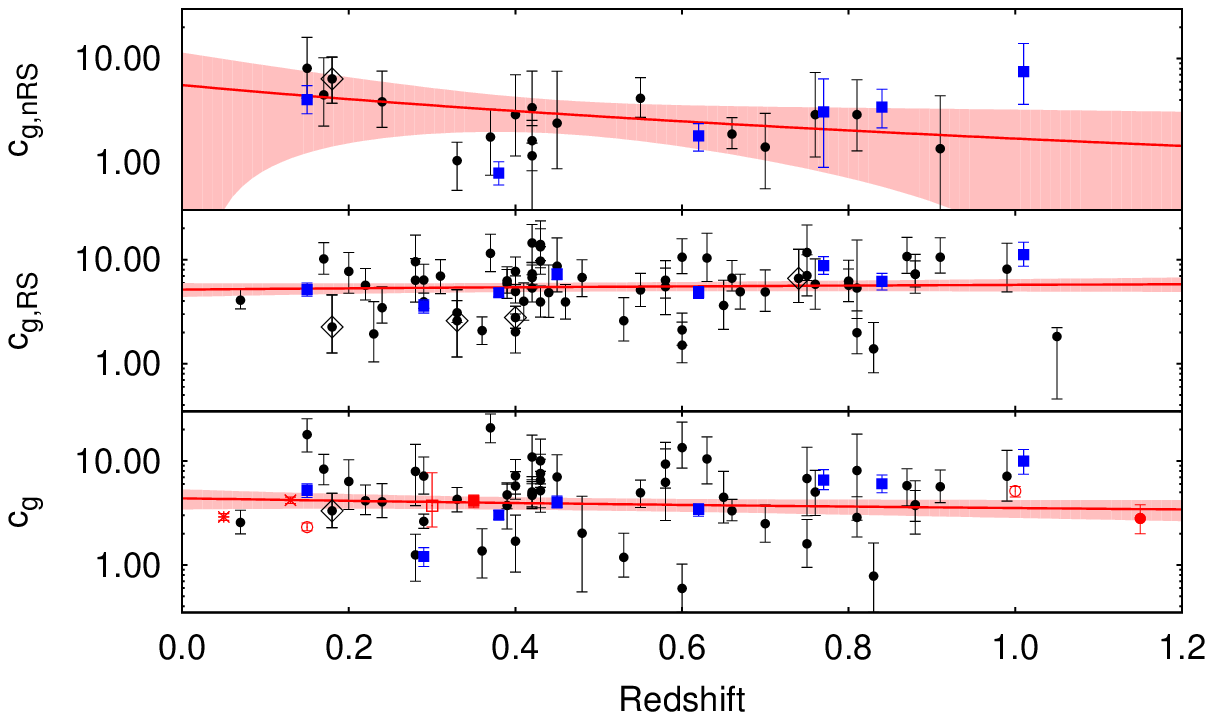}
  \vskip-0.5in
  \caption[]{Redshift trend of the concentration parameter for the non-RS blue (top), RS (middle) and full (bottom) populations. The black points (blue) show the best fit NFW concentration for individual clusters (stacks).  The line marks the best fit redshift trend and 1$\sigma$ region (see Table~\ref{tab:fitvalues}).  The characteristic concentrations for the non-RS, RS and full populations are $1.59 \pm 0.53$, $5.47 \pm 0.53$ and $3.89\pm 0.52$, respectively.  There are no statistically significant mass or redshift trends.  Considerable cluster to cluster scatter at the 55\% to 38\% level is apparent.  The red points show various published results that are described in the text (Section~\ref{sec:concentrationresults}).  In addition, we mark the clusters close to the Large Magellanic Cloud ($\delta<-63^\circ$) as black open diamonds.}
\label{fig:evo_c}
\vskip-0.25cm
\end{figure*}

\subsection{Radial Distribution of Galaxies}
\label{sec:profile}
We study the radial profile of the galaxy number density because it is a fundamental property of the population, but we also need the radial profile to enable a statistical correction for the cluster galaxies that are projected onto the cluster virial region but actually lie outside the virial sphere in front of or behind the cluster.  In the following we describe the profile fitting method and testing (Section~\ref{sec:NFWfitting}), the method for constraining trends in mass and redshift (Section~\ref{sec:trendfitting}) and our results on the concentration (Section~\ref{sec:concentrationresults}).

\subsubsection{Fitting Projected NFW Profiles}
\label{sec:NFWfitting}

To construct the radial profile we measure the number of galaxies lying within annuli centered on the cluster.  It has previously been shown that the offsets between SZE centers and BCG positions \citep{song12b} are consistent with the offsets measured between X-ray centers and BCGs \citep{lin04b}, but the SZE positional measurement uncertainties are large compared to the BCG positional uncertainties.   Thus, for this analysis we adopt the BCG position as the cluster center.  BCGs are selected manually through visual inspection of the pseudo-color images.  If there is no clear, centrally located BCG we adopt the brightest galaxy within $0.5*R_{200}$ that has a color within 0.22~mag of the RS color at that redshift and is located closest to the SZE center.  In eight cases, this BCG definition leads to the selection of what is clearly a bright foreground galaxy, and in these cases we exclude those galaxies and select a fainter BCG candidate.

The radial profile extends to between $\sim 4R_{200}$ and $\sim 14R_{200}$, given the $1^\circ\times1^\circ$ or $2^\circ\times2^\circ$ tiles we prepare for each cluster.  Thus, in all cases it includes a background dominated region.  We correct the individual profiles for bright stars that contaminate the cluster and background areas. For each profile annulus we calculate an effective area by subtracting off the star areas that are contaminating the bin.  Bright stars are selected from the 2MASS survey using a magnitude cut of $J<13.5$. We use an empirically calibrated relation between the J-band magnitude and the masking radius of the star to exclude spurious objects.  For the profile analysis we use galaxies that are brighter than $\sim m_*+2$ in the band redward of the 4000~\AA\ break, except again in the highest redshift bins where our imaging depth does not allow analysis to the full depth and in the highest redshift bin where $z$ band contains the 4000~\AA\ break. All profiles are completeness corrected as described in Section~\ref{sec:completeness}.

We fit these profiles to the NFW \citep{navarro97} density profile with the concentration as one of the free parameters. The three dimensional NFW profile is given as:
\begin{equation}
	\frac{\rho(r)}{\rho_\mathrm{c}}= \frac{\delta_c}{(r/r_\mathrm{s})(1+r/r_\mathrm{s})^2} 
\end{equation}
where $\rho_\mathrm{c} = 3H_0^2 / 8 \pi G$ denotes the critical density of the Universe, $\delta_\mathrm{c}$ is a characteristic density contrast and $r_\mathrm{s}$ is the typical profile scale radius.  The concentration parameter for the NFW profile is defined as $c=R_{200}/r_\mathrm{s}$.  In our application we measure the galaxy surface density profile, and therefore we use the projected and integral projected versions of the NFW model.  

Our model profile is the superposition of the cluster profile $\Sigma_{\mathrm{cl}}$ and a constant background $\Sigma_{\mathrm{back}}$:
\begin{equation}
	\Sigma (x) = \Sigma_{\mathrm{cl}} + \Sigma_{\mathrm{back}}
\end{equation}
Consequently the formula for the projected NFW profile has three free parameters: the normalization, the constant background and the scale radius $r_\mathrm{s}$ (or, equivalently, the concentration $c$).  We follow \citet{lin04a} in adopting the integrated number of galaxies within \Rtwohundred\ as the normalization.  This avoids the parameter degeneracy between the concentration and the central density and results in improved constraints on the concentration.   To fit the profile we use the \citet{cash79} statistic and the observed and predicted total counts per bin, rather than the number density per bin. To determine the predicted counts per bin we evaluate the model as $N(R_{\mathrm{up}})-N(R_{\mathrm{low}})$, where $N(R_{\mathrm{up}})$ is the integrated number of galaxies inside a projected cylinder with an outer radius of $R_{\mathrm{up}}$, and $N(R_{\mathrm{low}})$ is the equivalent inside a cylinder of Radius $R_{\mathrm{low}}$.  This avoids biases introduced by binning in the inner regions of the profile where it is changing rapidly with radius.  

Due to the star-masking, we are missing galaxies inside the radial bin and so we correct the model number of galaxies with the ratio $A_{\mathrm{eff}}/A_{\mathrm{true}}$, where $A_{\mathrm{eff}}$ is the effective bin area after star-masking, and $A_{\mathrm{true}}= \pi(R_{\mathrm{up}}^2-R_{\mathrm{low}}^2)$ represents the geometric bin area. As the measured number inside a radial bin is the sum of cluster galaxies and background galaxies, we add the background contribution to the model with $\Sigma_{\mathrm{back}}*A_{\mathrm{eff}}$. In summary, within each annulus we measure the number of observed galaxies, calculate the effective area and fit a model that reads
\begin{equation}
	N(r)= (N(R_{\mathrm{up}})-N(R_{\mathrm{low}}))\frac{A_{\mathrm{eff}}}{A_{\mathrm{true}}}+ \Sigma_{\mathrm{back}}A_{\mathrm{eff}}
\end{equation}
From integrating the surface density, $N(R_{\mathrm{up}})$ becomes
\begin{equation}
N(R_{up}) = 4\pi \rho_\mathrm{s} r_\mathrm{s}^3 * g(x)
\end{equation}
\begin{displaymath}
g(x)=
\begin{cases}
  \frac{2}{\sqrt{(x)^2-1}}\mathrm{arctan}\sqrt{\frac{x-1}{x+1}} + \mathrm{ln}(\frac{x}{2}) & \text{if } x > 1, \\
  \frac{2}{\sqrt{1-(x)^2}}\mathrm{arctanh}\sqrt{\frac{1-x}{1+x}} + \mathrm{ln}(\frac{x}{2}) & \text{if } x < 1\\
  1+ \mathrm{ln}(\frac{x}{2}) & \text{if } x = 1
\end{cases}
\end{displaymath}
where $x=cR_{\mathrm{up}}$. $N(R_{\mathrm{low}})$ is calculated correspondingly \citep{bartelmann96}.   The profile fitting is done with the Markov chain Monte Carlo (MCMC) Ensemble sampler from \citet{mackey13}. Because the error distribution for the concentration parameter is closer to lognormal than to normal, we fit for $\mathrm{ln}(c)$.  Figure~\ref{fig:constrain} contains an example for the profile parameter constraints for a stacked cluster profile containing low redshift systems.

We test the profile generation and the fitting procedure on a sample of mock catalogs with a concentration of $c=5$. We build a large mock catalog with 500,000 galaxies within a projected region extending to 10$R_{200}$ to first ensure our code returns unbiased results in the high signal to noise limit.  We test first with no background and then create a new mock catalog with the background adjusted to be comparable to what we see in the real data. We then draw 100 individual realizations with different numbers of cluster galaxies-- that is 100, 500, 1000, 2000 and 4000 galaxies-- to test behavior in the limit where the Poisson noise is important. We find that even in the low signal-to-noise regime with just 100 cluster galaxies (which is typical also for the SPT sample we probe here), we can fully recover the input concentration with an inverse variance weighted mean of  $4.95 \pm 0.16$. The normalization is also recovered to within the $1\sigma$ statistical uncertainty.   

With our fitting approach the radial bins can be infinitesimally small. For our application we use bins of 0.02 inside \Rtwohundred, and increase the bin size outside this radius.  Our tests show that an overestimation of the background leads to overestimated concentrations, and therefore it is very important to fit a region extending to large enough radius to constrain both the background and the cluster model.  In particular, we find that it is important to have data extending to $\sim4R_{200}$, and so we have ensured that our fits include data out to this radius for all clusters in the sample.  Through these tests we have demonstrated that our profile fitting code and approach are unbiased to within the (small) statistical uncertainties in our tests.

\subsubsection{Fitting Trends in Mass and Redshift}
\label{sec:trendfitting}

We fit a simple power law to the log of the concentration parameter of the individual cluster profiles simultaneously in mass and redshift.  Because this approach is used in the other observables examined below, we define the relation here for a generic observable $O(M_{200},z)$ as:
\begin{equation}
  O(M_{200},z)= A \left( \frac{M_{200}}{M_\mathrm{piv}} \right)^{B} \left(\frac{1+z}{1+z_\mathrm{piv}}\right)^C
  \label{eq:fit}
\end{equation}
where $A$ is the normalization, $B$ is the mass power law index and $C$ is the redshift power law index. We choose a mass pivot point $M_{\mathrm{piv}}=6\times10^{14} M_{\odot}$, which is the median mass of our sample, and a redshift pivot point $z_\mathrm{piv}=0.46$, which is the median redshift of our sample.  

In addition to these three parameters we constrain the intrinsic scatter $\sigma_{int}$ of these relations.  With this intrinsic scatter the uncertainty on a given parameter measurement becomes the quadrature addition of the intrinsic and measurement uncertainties.  We iterate our fit, adjusting the intrinsic scatter until the reduced $\chi^2$ for the fit approaches 1.0.  For all cases except for the RS fraction the intrinsic scatter is evaluated as the fractional scatter.  These best fit parameters and estimates of the intrinsic scatter for all observables considered are listed in Table~\ref{tab:fitvalues}. 

\begin{figure*}
\vskip-1.0cm
  \includegraphics [width=1.0\textwidth]{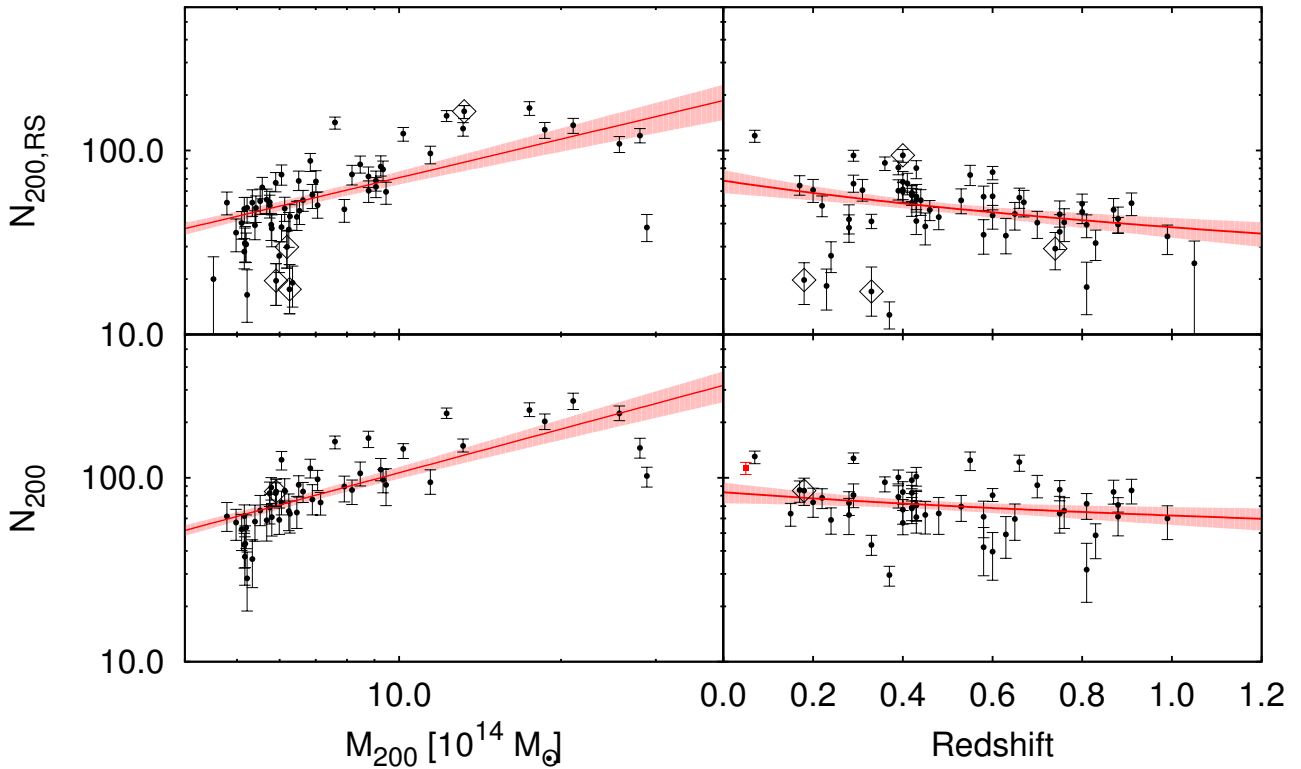}
  \vskip-0.50in
  \caption [Trends of the Number of galaxies with mass and redshift]{The HON from measurements of $N_{200}$ from the galaxy radial profiles is plotted versus mass (left panels), and the HON normalized to a mass of $6\times10^{14} M_{\odot}$ is plotted versus redshift (right panels).  In both cases the full (RS) population appears on bottom (top), and both populations have $N_\mathrm{g}\propto M^B$ where $B<1$ (see Table~\ref{tab:fitvalues}). We find no clear evidence of a redshift trend in the full population, but the RS population shows a 2.5$\sigma$ trend for falling HON with redshift.  The 1$\sigma$ region (Equation~\ref{eq:fitsigma}) for each fit is plotted in red.  The clusters near the LMC ($\delta<-63^\circ$) are more contaminated by stars and are marked with black open diamonds.  
  \label{fig:N200}}
  \vskip-0cm
\end{figure*}

Using the best fit parameters and parameter uncertainties we use Gaussian error propagation to estimate the uncertainties on the best fit as:
\begin{align}
\label{eq:fitsigma}
\frac{\sigma^2_\mathrm{O}}{O^2}(M_{200},z)=& \frac{\sigma^2_\mathrm{A}}{A^2}+\left(\log{\frac{M_{200}}{M_{\mathrm{piv}}}}\right)^2\sigma^2_\mathrm{B} \\
&+\left(\log{\frac{(1+z)}{(1+z_{\mathrm{piv}})}}\right)^2\sigma^2_\mathrm{C} \nonumber
\end{align}
where the parameter errors are extracted from the covariance matrix of the fit and off-diagonal terms are ignored.  We use this uncertainty to define a confidence region about the best fit relations (see Table~\ref{tab:fitvalues}) when they are presented in Figures \ref{fig:evo_c}, \ref{fig:N200} and \ref{fig:redRP}.

\subsubsection{Observed Galaxy Radial Profile Concentrations}
\label{sec:concentrationresults}

We first present the radial profile of the stacked clusters for the full, the RS and the blue non-RS populations to provide for a comparison of the three populations. For the stacks we sum the number of galaxies and the corresponding areas within each radial bin. This sum is then averaged by the number of contributing clusters inside the bins.   The blue non-RS population is selected to have a color that is blueward from the cluster RS and each galaxy is weighted with $1-P_{\mathrm{i,j}}$ as in Equation \ref{eq:red}.  These stacks are shown for the full cluster sample in Figure~\ref{fig:RP_stack} and for the 8 different redshift bins in Figure~\ref{fig:RP_stack_bins} with black, red and blue points and best fit models for the full, RS and blue non-RS populations, respectively.  All individual profiles extend out to $4R_{200}$. 

We find that the full population is less concentrated with $c_\mathrm{g} = 3.59 ^{+0.20} _{-0.18}$ compared to the RS population with $c_{\mathrm{g,RS}} = 5.37 ^{+0.27} _{-0.24}$.  The blue, while clearly having a density that increases toward the cluster center, is even less concentrated with $c_{\mathrm{g,non-RS}} = 1.38 ^{+0.21} _{-0.19}$ (see Figure~\ref{fig:RP_stack}). The same general picture arises in each redshift bin as seen in Figure~\ref{fig:RP_stack_bins}. Note also that the blue non-RS background is higher than the RS background, making the study of the blue non-RS population more challenging. In redshift bins 2 and 4 the concentration of the blue non-RS population was unconstrained due to large scatter, and therefore we show no model.  Note that our data provide no convincing evidence of a departure from the NFW model out to 4$R_{200}$ in any population; given the expected amplitude of the deviations due to neighboring halos in the dark matter \citep[e.g.][]{hilbert2010} and the scale of our measurement uncertainties, this is not surprising.

\begin{center}
\begin{table*}
   \caption{Properties of the stacked profiles.  We list the bin number, redshift and mean followed by the concentration and $N_{200}$ for the full, RS and non-RS blue stacks.  The last two columns contain the RS fraction and the intrinsic scatter of the RS.}\label{tab:stack_value}
    \begin{tabular}{lcccccccccc}
\hline \hline  
 & & $\left< M_{200}\right >$  \\ 
 \# & $z$ &  [$10^{14}M_\odot$] & $c_{\mathrm{g,st}}$ & $N_{200,\mathrm{st}}$ & $c_{\mathrm{g,RS,st}}$ & $N_{200,\mathrm{RS,st}}$ & $c_{\mathrm{g,nRS,st}}$ & $N_{200,\mathrm{nRS,st}}$ &  $\sigma_{\mathrm{int,g-r}}$\\ \hline \hline                             
 1 & 0.15 &   6.10   &$5.19^{+0.78}_{-0.75}$&  $106.04^{+6.40}_{-6.17}$   & $5.16^{+0.79}_{-0.69}$&  $76.72^{+3.77}_{-3.67}$  &  $4.03^{+1.46}_{-1.08}$&  $29.61^{+4.07}_{-3.89}$  & $0.033\pm0.004$ \\[3pt]
 2 & 0.29 &   9.97 &   $1.02^{+0.22}_{-0.20}$&  $157.90^{+9.45}_{-8.56}$  & $3.35^{+0.50}_{-0.47}$&  $79.50^{+3.80}_{-4.45}$  &  - & -   & $0.026\pm0.006$ \\[3pt]
 3 & 0.38 &   10.40 &$2.52^{+0.26}_{-0.22}$&  $210.17^{+7.54}_{-7.12}$  & $4.48^{+0.47}_{-0.40}$&  $121.77^{+3.73}_{-4.13}$ &  $0.79^{+0.23}_{-0.18}$&  $91.43^{+6.24}_{-5.94}$  &  $0.040\pm0.003$ \\[3pt]
 4 & 0.45 &   8.71 &$3.54^{+0.53}_{-0.41}$&  $135.58^{+6.57}_{-6.67}$    &$6.92^{+0.85}_{-0.79}$&  $80.23^{+3.39}_{-3.52}$    &  - & -   & $0.026\pm0.006$ \\[3pt]
 5 & 0.62 &   6.12 &$3.39^{+0.46}_{-0.47}$&  $113.28^{+5.28}_{-5.88}$  & $4.82^{+0.70}_{-0.58}$&  $71.90^{+3.24}_{-3.24}$   &  $1.80^{+0.56}_{-0.52}$&  $40.25^{+4.77}_{-4.19}$ & $0.037\pm0.002$ \\[3pt]
 6 & 0.77 &   5.78 &$6.62^{+1.73}_{-1.25}$&  $70.04^{+6.91}_{-7.17}$   &  $8.81^{+1.94}_{-1.53}$&  $45.25^{+3.40}_{-3.43}$  &  $3.07^{+3.30}_{-2.18}$&  $22.04^{+8.22}_{-6.82}$ & $0.036\pm0.029$ \\[3pt]
 7 & 0.84 &   8.34 &$5.43^{+1.16}_{-0.99}$&  $75.23^{+6.75}_{-6.70}$   & $5.93^{+1.17}_{-1.03}$&  $49.44^{+3.54}_{-3.57}$ &  $3.41^{+1.65}_{-1.25}$&  $26.11^{+5.10}_{-5.19}$ & $0.033\pm0.007$ \\[3pt]
 8 & 1.01 &   5.37 &$10.30^{+3.06}_{-2.55}$&  $39.22^{+4.21}_{-4.91}$    & $11.40^{+3.58}_{-2.61}$&  $24.36^{+2.20}_{-2.51}$  & $7.48^{+6.50}_{-3.85}$&  $13.69^{+3.44}_{-4.13}$  & $0.070\pm0.010$ \\ 
 \hline\hline    \end{tabular}
    \end{table*}  
\end{center}

The best fit parameter values for the cluster stacks are shown in Table~\ref{tab:stack_value}, and the best fit mass and redshift trends (Equation~\ref{eq:fit}) of the concentration in the stacked sample is presented in Table~\ref{tab:fitvalues}.   The characteristic value of the concentration for the full population at the pivot mass $M_{\mathrm{piv}}=6\times10^{14}M_\odot$ and the pivot redshift $z_{\mathrm{piv}}=0.46$ is $c_{\mathrm{g,st}}=3.19\pm0.64$, and there are no statistically significant redshift trends, although there is 2.2$\sigma$ and 2.4$\sigma$ evidence for higher concentrations at higher redshift in the stacked full and RS populations, respectively.  For the RS population the characteristic value is $c_{\mathrm{g,RS,st}}=5.33\pm0.53$, also with no statistically significant trends. The blue non-RS population is even less concentrated with $c_{\mathrm{g,nRS,st}}=1.59\pm0.53$. Similar to the other populations there is no mass and redshift trend for the blue non-RS population. These fits to the concentrations from the stacks in the eight redshift bins are consistent with the stack of the entire cluster ensemble shown in Figure~\ref{fig:RP_stack}.

We also measure concentrations for individual clusters, and these are listed in Table~\ref{tab:properties}.  Fitting these individual cluster concentrations to the power law trends in mass and redshift, we find that the characteristic concentration for the full population is $c_\mathrm{g}=3.89\pm0.52$ and that there is considerable cluster to cluster intrinsic scatter of 55\%.  Consistent with the results from the stacks, there is no statistically significant evidence for a redshift or mass trend (see Table~\ref{tab:fitvalues}), although there is a preference at 1.8$\sigma$ for more massive systems to have lower concentration.  

Figure~\ref{fig:evo_c} (lower panel) contains a plot of the galaxy population concentration for each cluster (black points) and for each stack (blue squares) versus redshift.  Measurements from the literature are plotted in red with different point styles:   \citet{capozzi12} (red filled circle), \citet{popesso06} (red cross), \citet{lin04a} (red star), \citet{carlberg97} (red open square), \citet{muzzin07a} (red filled square) and \citet{vdB15} (red open circles).  There is good agreement among these previous results and our own. The considerable cluster to cluster variation in concentration of 55\% (listed for all fits in last column of Table~\ref{tab:fitvalues}) is clearly apparent in this figure.

We use the same technique to measure the concentrations for the RS galaxy population, where we assign each galaxy a probability of being a RS member as described in Equation \ref{eq:red}. The concentration trend with redshift is shown in Figure~\ref{fig:evo_c} (middle panel).  The characteristic concentration at our pivot mass and redshift is $c_{\mathrm{g,RS}}=5.47\pm0.53$. Consistent with the results from the stacks, we do not find statistically significant mass and redshift trends.  The somewhat lower cluster to cluster scatter of 38\% is also apparent in the figure.  

We also study the concentrations of the blue non-RS population, although for more than two thirds of the clusters the blue non-RS concentrations are unconstrained.  The characteristic value for this subset of cluster population is $c_{\mathrm{g,nRS}}\sim3.35\pm0.38$ (see Figure~\ref{fig:evo_c} top panel), which is higher than the values seen in the stacks created within redshift bins.  We note that an agreement  between the stacks and the individual profiles is not guaranteed because the stacks contain all clusters in the sample and represent a galaxy number weighted average, whereas we report concentrations for the individual profiles only in the cases where the fit converges.  Because it requires higher signal to noise to constrain a fit to a galaxy distribution with lower concentration, it is preferentially the low concentrations that are unconstrained.  This may lead to a tendency for the stacks, which represent the full population, to show lower concentrations than the fit to the individual measurements.  Because such a small fraction of individual clusters lead to concentration measurements, we believe that in this case the stacks provide a more robust result.


In summary, our data show that the red, early-type galaxies tend to have a more concentrated distribution than the blue non-RS and full populations, which is consistent with previous analyses where a higher concentration is seen in the red galaxy population \citep[e.g.,][]{goto04}.  Spectroscopic studies of individual clusters also support this picture that there is a flatter distribution of emission line galaxies centered on the cluster whose virial region is strongly marked by a more concentrated distribution of absorption line systems \citep[e.g.,][]{mohr96}.  Our analysis clearly indicates that  there is a clustered blue population with lower concentration associated with the clusters over the full redshift range.  Moreover, our sample allows for a consistent study of the galaxy populations of these systems out to $z\sim1.1$.  Rather than indicating a clear redshift trend, the most salient feature of the concentrations for the sample is the large cluster to cluster scatter.

\subsection{Halo Occupation Number}
\label{sec:n200}
The halo occupation distribution (HON) describes the relation between galaxies and dark matter at the level of individual dark matter halos  \citep{zheng05}, providing insight into how baryonic matter is distributed within each of the dark matter halos.  A key feature here is the relation between the halo occupation number and $M_{200,\mathrm{c}}$.  A simple prediction based on galaxy formation efficiency is that the number of galaxies formed is proportional to the baryonic mass within the halo. The HON-mass relation becomes $N_{\mathrm{g}} \sim M^{B}$ with $B >1$ if galaxy formation is more efficient in the more massive haloes, or $B<1$ the other way around. Reasons for $B<1$ include that the gas, which is heated through the collapse of the halo, may not effectively cool and collapse into galaxies in the higher mass halos.  In addition, dynamical friction and tidal stripping can have an impact on the HON, which is typically measured above a magnitude threshold \citep{lin04a,rines04}. 

Here we calculate the HON directly from the normalization $N_{200}$ of the individual fits to the radial profile of the cluster galaxies-- both the full sample and the RS selected sample. We correct the measured $N_{200}$ inside the virial cylinder to the virial sphere. Therefore we calculate the correction factor $f_{c}$ from the integration of the projected surface number density inside a cylinder and sphere as
\begin{equation}
	f_{\mathrm{c}}=\frac{N_{\mathrm{cyl}}(R_{200})}{N_{\mathrm{sph}}(R_{200})}
	\label{eq:C}
\end{equation}
These integrals are simply a function of the concentration $c$. For a concentration $c=3$, for example, the correction factor  is $f_{\mathrm{c}}\sim$1.3.  For each studied cluster we use the measured concentration to correct $N_{200}$.  In addition, in the highest redshift bins where the completeness does not extend completely to $m_*+2$ we use the faint end slope measured for the individual cluster or the stack, as appropriate, to apply a correction to the $N_{200}$.  In this way, all $N_{200}$ measurements refer to the population within the virial sphere brighter than $m_*+2$.

Figure~\ref{fig:N200} is a plot of the measured $N_{200}$ for each cluster versus mass (left) and redshift (right) for the full population (bottom) and RS population (top).  The best fit power law parameters (Equation~\ref{eq:fit}) describing these data appear in Table~\ref{tab:fitvalues}. The characteristic $N_{200}$ at our pivot redshift $z_\mathrm{piv}=0.46$ and mass $M_\mathrm{piv}=6\times10^{14}\,M_\odot$ is $71.1\pm3.9$ for the full population and $49.8\pm2.9$ for the RS population.  This is an indication that the red fraction at the pivot mass and redshift to $m_*+2$ is $(70 \pm 7)$\% for the galaxy population within the $R_{200}$ virial sphere.  

The mass trend for the full population $N_{200}\propto M^{(0.79\pm0.10)}$ is consistent with the RS population, where we find $N_{200,RS}\propto M^{(0.70\pm0.11)}$. The full population shows no statistically significant evidence of redshift variation $N_{200}\propto(1+z)^{-0.42\pm0.31}$, but the RS population varies with redshift as $N_{200}\propto(1+z)^{-0.84\pm0.34}$, a trend that is statistically significant at 2.5$\sigma$.  This difference in behavior is suggestive of an increase in RS fraction over cosmic time.

As with the other power law fits, we extract constraints on the variation of the $N_{200}$ from cluster to cluster for both the RS and full populations.  We find a 31\% intrinsic scatter in $N_{200}$ around the best fit relation for the full population, and a 37\% intrinsic scatter for the RS population.  This evidence of scatter is corrected for the Poisson sampling noise and background correction uncertainties that are included in the $N_{200}$ measurements.  Thus, both the concentration and number of galaxies vary significantly from cluster to cluster of a given mass and redshift.  Our estimates of intrinsic scatter of $N_{200}$ are larger by a factor of 1.5 to 2 in comparison to recent measurements of the intrinsic scatter in the richness measure $\lambda$ of a sample of similar SZE selected clusters \citep{saro15}.  The richness measure $\lambda$ is extracted from a somewhat smaller portion of the cluster virial region.


We do not fit mass and redshift trends for the stacked clusters (but results of all stack measurements are listed in Table~\ref{tab:stack_value}), because the direct stacking approach we use leads to a noisy and biased estimate of $N_{200}$.  Specifically, within the redshift bins we create a stack by summing the individual cluster profiles.  Within the limit of similar backgrounds, this leads to an $N_{200}$ weighted stack, where naturally those clusters with the most contributed galaxies have the largest impact on the stack.  This leads to a noisy and biased estimator of $N_{200}$ that is approximately $\left<N^2_{200}\right>/\left<N_{200}\right>$.  This is a generic problem with stacking that is worse for our sample, because we have a large mass range in each redshift bin and too few clusters in each bin to subdivide into mass bins to create more similar subsamples.  For the concentration $c$ this weighting is not as problematic because we see no mass dependence in the concentrations, and in the fit the parameters concentration $c_\mathrm{g}$ and $N_{200}$ are only weakly correlated.

We find agreement between our measured mass trend in the full population and previously published results: $B=0.87 \pm 0.04$ \citep{lin04a}, $B=0.70 \pm 0.09$ \citep{rines04} and $B=0.92 \pm 0.03$ \citep{popesso07b}.  Comparing the normalization to the local sample at $z\sim0.05$ of \citet{lin04a}, we first correct the $m_*+3$ values used in the \citet{lin04a} analysis to the $m_*+2$ used in our analysis, adopting their faint end slope $\alpha=-1.1$.  Our normalization is lower than that observed by \citet{lin04a} with a low redshift $z<0.2$ sample (see Figure~\ref{fig:N200} lower right panel), where we have few clusters in our sample.


\begin{figure}
\vskip-0.05cm
  \includegraphics [width=0.48\textwidth]{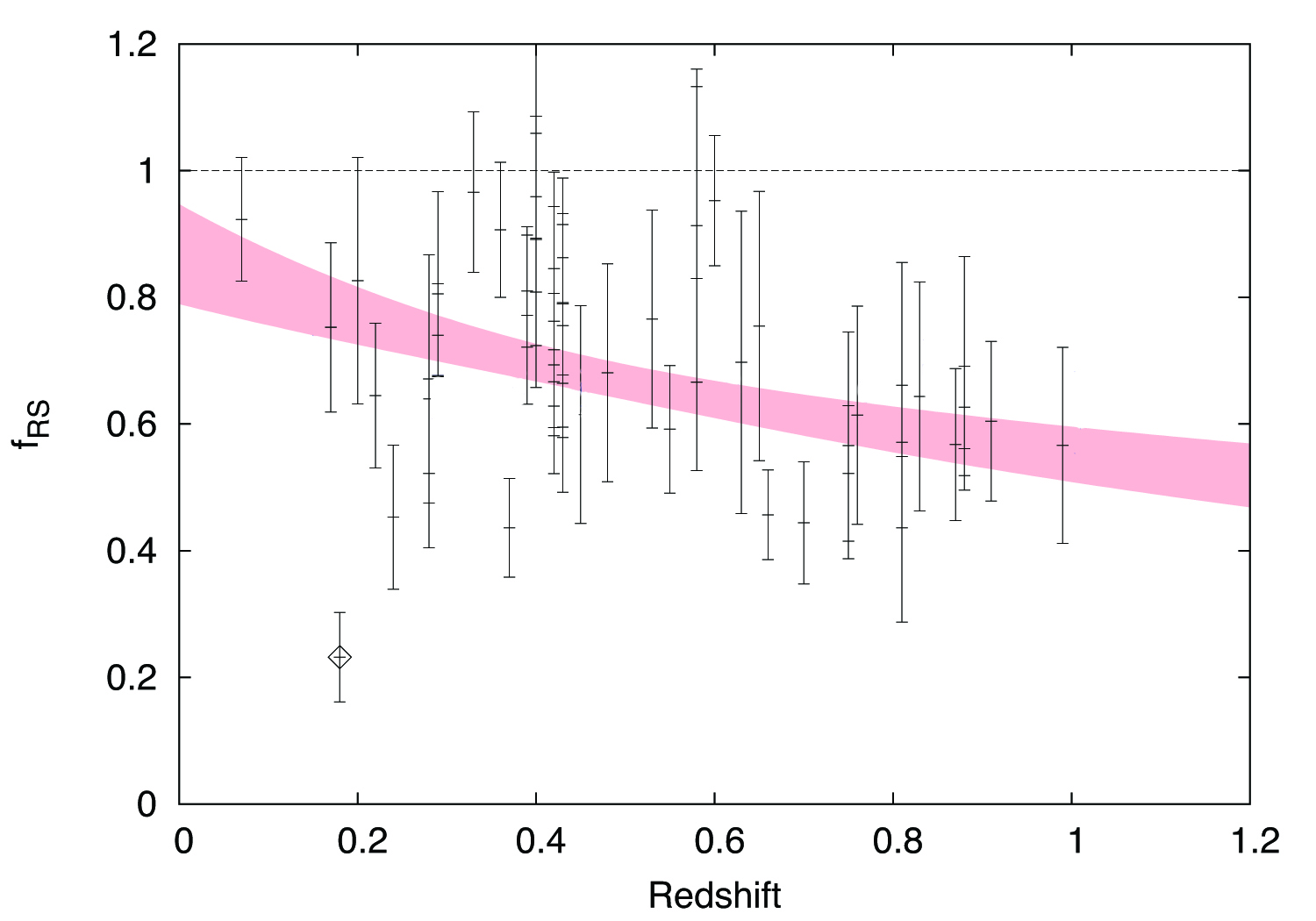}
  \vskip-0.20cm
  \caption [Evolution of the Red Fraction with redshift]{Red sequence fraction within the virial sphere of radius \Rtwohundred\ as a function of redshift. The individual $f_\mathrm{RS}$ measurements are in black, and best fit trend is shown in pink. The typical RS fraction is $68\pm3$\% at the pivot redshift $z=0.46$ and mass $M=6\times10^{14}M_\odot$ with intrinsic variation at the 14\% level.  The best fit redshift trend $f_\mathrm{RS}\propto(1+z)^{-0.65\pm0.21}$ has characteristic values around 55\% at $z=1$ and 80\% at $z=0.1$.  One LMC cluster appears marked with an open diamond.
  \label{fig:redRP}}
  \vskip-0cm
\end{figure}

\subsection{Red Sequence Fraction}
\label{sec:redfraction}
For studying the evolution of the RS fraction $f_{\mathrm{RS}}$, we look at the galaxy population brighter than $\sim m_*+2$ that lies within the cluster virial sphere ($r<R_{200}$).  Specifically, we calculate the ratio $f_{\mathrm{RS}}=N_{200,\mathrm{RS}}/N_{200}$, where the $N_{200}$ are extracted for each cluster during the radial profile fitting (see Section~\ref{sec:NFWfitting}).  Note here that these measurements are corrected to the virial sphere using the appropriate individual concentration measurements for each population as clarified in Equation~\ref{eq:C}.  As described in Section~\ref{sec:RSselection}, the RS galaxies are selected with a probabilistic approach based on the stacked color distributions in 8 redshift bins.  We derive $f_{\mathrm{RS}}$ for the individual clusters, but as discussed in the previous section, the stacks for our sample are not good estimators for $N_{200}$, and therefore we do not use them to estimate $f_{\mathrm{RS}}$. Note that given measurement noise, it is possible for the red fraction estimate for a single cluster to scatter above 1.

A plot of $f_{\mathrm{RS}}$ versus redshift appears in Figure~\ref{fig:redRP}. The red area represents the best fit function to the individual cluster measurements, and the best fit parameters are given in Table~\ref{tab:fitvalues}.  We find that at the pivot mass and redshift the RS fraction is  $f_{\mathrm{RS}}=(68\pm3)$\%, which is in agreement with the characteristic values of the $N_{200}$ for the full and RS populations presented in Section~\ref{sec:n200}.  The RS fraction is decreasing with redshift $f_{\mathrm{RS}}\propto(1+z)^{-0.65\pm0.21}$ in the individual measurements with $3\sigma$ significance. We find a decrease in RS fraction from $\sim 80\%$ at $z\sim0.1$ to $\sim 55\%$ at $z\sim1$. 

Our result is in reasonably good agreement with previous studies at low redshifts.  Studies of optically selected systems in SDSS at low redshift indicate red fractions within a projected $R_{200}$ for clusters with $N_{200}>50$ of $\sim$82\% at $0.1<z<0.25$ and $\sim$78\% at $0.25<z<0.35$ \citep{hansen09}.  These are projected rather than virial sphere RS fractions, and the RS fraction is defined with a flat color bin rather than, as in our case, a probabilistic approach supported by the color distribution of the data.  A separate analysis of SDSS clusters that also relied upon optical mass indicators yielded RS fractions within a fixed metric aperture of $\sim$1~Mpc of between $\sim$90\% and $\sim$70\% in the redshift range $z=0.12$ to $z=0.4$ \citep{zhang13}.  Both results seem to be in good agreement with an earlier DPOSS-II study \citep{margoniner01}.   At higher redshift it is more difficult to compare published results to ours, because many of those RS fraction measurements have been made only in the central regions of the cluster, have been made using a stellar mass cut rather than the $m_*+2$ magnitude cut we adopt and have relied on different definitions of the RS than ours.  Nevertheless, we find that \cite{loh08} present a study of clusters selected in the Red Cluster Sequence survey to $m_*+1.5$ that shows falling RS fraction with redshift, ending with $\sim0.55$ in their highest redshift bin ($0.85\le z\le0.90$) in good agreement with the trend we constrain over a somewhat different magnitude range with our SZE selected cluster sample.


\begin{center}
\begin{table}
   \caption{Mass and redshift trends for the individual cluster properties using Equation~\ref{eq:fit} with $M_\mathrm{piv}=6\times10^{14}M_\odot$, $z_\mathrm{piv}=0.46$. The rows contain the best fit parameters for the galaxy concentration $c$, the number of galaxies $N_{200}$, the RS fraction and intrinsic width of the RS.  All measurements are made within $R_{200}$.  Where possible, the results are shown for the full, RS and non-RS blue populations.  The columns contain the observable, as well as the best fit normalization $A$, mass slope $B$ and redshift slope $C$. The last column shows the intrinsic fractional scatter about the best fit relation.}\label{tab:fitvalues}
    \begin{tabular}{lrrrc}
    \hline \hline   
 Obs & \multicolumn{1}{c}{A} & \multicolumn{1}{c}{B} & \multicolumn{1}{c}{C} &  $\sigma_{int}$\\ \hline \hline                             
$c_\mathrm{g}$                     &$3.89 \pm 0.52$  & $-0.32\pm0.18$ &$-0.31 \pm 0.45 $ & 0.55\\[3pt]
$c_\mathrm{g,RS}$             &$5.47 \pm 0.53$  & $-0.01\pm0.10$ &$0.15 \pm 0.30 $ & 0.38\\[3pt]
$c_\mathrm{g,nRS}$             &$3.35 \pm 0.38$  & $-0.89\pm0.28$ &$-1.85 \pm 0.55 $ & 0.00\\[3pt]
$N_{200}$	            &$71.1 \pm 3.9$  & $0.79\pm0.10$ &$-0.42 \pm 0.31 $ & 0.31\\[3pt]
$N_{200,\mathrm{RS}}$            &$49.8 \pm 2.9$  & $0.70\pm0.11$ &$-0.84 \pm 0.34 $ & 0.37\\[3pt]
$f_\mathrm{RS}$				&$0.68 \pm 0.03$  & $-0.10\pm0.06$ &$-0.65 \pm 0.21 $ & 0.14\\[3pt]
 $\sigma_{\mathrm{int,RS}}$   &$0.036 \pm 0.002$  & \multicolumn{1}{c}{-}			 &$0.58 \pm 0.47 $ & -\\[3pt]



 \hline
    \end{tabular}
    \end{table}  
\end{center}


\section{Conclusions}
\label{sec:conclusions}
We present results from a study of 74 SZE selected clusters with redshifts extending to  $z \sim1.1$ from within the overlapping regions of the SPT and DES-SV survey areas. The combination of the deep DES data and this SPT selected cluster sample provides a unique opportunity to systematically study the galaxy population and its redshift and mass trends in high mass clusters over a wide range of redshift.  Each of these clusters has a robust mass estimate that derives from the cluster SZE detection significance and redshift;  these masses include corrections for SZE selection effects and account for the remaining cosmological uncertainties and unresolved systematics in the combined X-ray and velocity dispersion mass calibration dataset \citep{bocquet15}.  Our masses lie in the range of $4.3 \times 10^{14} M_{\odot}\le M_{200}\le 2.9\times 10^{15} M_{\odot}$. Within the cluster virial region \Rtwohundred, we study the width of the red sequence as well as the redshift and mass dependencies of the galaxy radial profiles, the HON and the RS fraction.

Our study of the radial distributions of galaxies in these clusters indicates that the blue, non-RS and full populations are consistent with NFW profiles of differing concentration out to 4$R_{200}$ (see Figure~\ref{fig:RP_stack}).  There is no compelling evidence for systematic density jumps within this region \citep[see discussion in][]{patej15}.  The blue, non-RS population has a much lower concentration than the RS population, leading to a red fraction that is a strong function of radius within the cluster.  The presence of a clustered, blue, non-RS population in all the cluster stacks reinforces the accepted scenario of ongoing infall from the field, which provides a continuous supply of star forming galaxies within the cluster virial region over the full redshift range of our sample.

In many of the clusters studied here there are enough galaxies to enable a measure of their individual radial galaxy distributions (see Section~\ref{sec:concentrationresults}).  These individual cluster measurements exhibit no statistically significant mass or redshift trend.   Our  measurements indicate significant variation from cluster to cluster both in the radial distributions of the galaxies and in the RS fractions, supporting a scenario where there is considerable stochasticity in the growth histories of galaxy clusters.  Moreover, the variation in radial distributions both of the full and RS populations undoubtedly also reflects the overall youth of these systems and the variety of merger states in which we observe them; this result echoes those long before established at low redshift using both the galaxy distributions, galaxy dynamics and X-ray morphologies \citep[e.g.,][]{geller82,dressler88,mohr95}.


The observed increase of RS fraction from $\sim$55\% at $z\sim1$ to $\sim$80\% at $z=0.1$ in the individual galaxy measurements (see Section~\ref{sec:redfraction}) constrains the timescale for the transition of a galaxy onto the RS \citep{mcgee09}.  A transition timescale of the order of 2 to 3~Gyr appears to provide a reasonable description of the trends in our sample, but further work is needed both on the observational and theoretical sides of this problem.  It must be emphasized that there have been studies where the redshift evolution of the red or blue fraction was either weaker or non-existent \citep[e.g.,][]{butcher84,smail98,depropris03,andreon06,haines09,raichoor12}, and various selection biases were considered as possible drivers for the trend.  Cluster samples that are uniformly selected using a property that does not involve their galaxy populations, that span a large redshift range, and that come with reasonably precise virial mass estimates are therefore desirable to provide maximal leverage for evolutionary studies.  In addition, precise and accurate cluster mass estimates are critical so that one is comparing similar portions of the cluster virial volume at different redshifts and masses.   From this perspective the SPT selected sample we study here offers some advantages to our study.  In addition, many current studies measure red fraction as a function of the stellar mass of the galaxies.  Given the DES band coverage $griz$ in our current dataset, this is not an approach we can apply over the full redshift range of our sample, and so we adopt a magnitude based selection.

Our study shows that the intrinsic rest frame $g-r$ RS color width has a characteristic value of $\sim0.03$ out to redshift $z\sim0.8$ and prefers a width of $\sim0.07$ at $z\sim1$ (see Section~\ref{sec:RSwidth}).  Our constraints at $z\sim1$ are consistent with those from recent HST-based studies \citep{mei09}, and are likely also consistent with what appears to be a higher scatter of $\sim0.074$ seen in an earlier study of low redshift clusters \citep[e.g.,][]{lopez-cruz04} once those earlier results are corrected for measurement uncertainties.  The RS width constrains the heterogeneity in age, metallicity and IMF at fixed galaxy luminosity of the old stellar populations that dominate RS galaxies.  More extensive analyses will be required to extract detailed population constraints from the RS width measurements from a well selected sample over this full redshift range, but the quite small intrinsic scatter out to $z\sim1$ underscores the homogeneity of the RS population.

Our sample provides no evidence for a significant redshift trend in $N_{200}$ for the full population, in agreement with most previous analyses \citep[][but see \cite{capozzi12}]{lin06,muzzin07,andreon14}.  Moreover, our study of the individual clusters indicates a characteristic intrinsic scatter of 31\% to 37\% from cluster to cluster at fixed mass and redshift.  This scatter is in addition to the Poisson noise and measurement uncertainties in $N_{200}$.  We observe the same mass trend for $N_{200}$ seen in earlier studies of local cluster samples \citep{lin04a,rines04}.  This indicates that there are fewer galaxies per unit mass in high mass clusters than in low mass clusters over the full redshift range extending to $z\sim1.1$.  At first glance this result seems strange-- suggesting that massive clusters do not reflect the properties of their lower mass building blocks.  However, within a hierarchical structure formation scenario where the more massive the cluster the larger the fraction of galaxies that have fallen in directly from the field-- bypassing the group environment \citep[e.g.,][]{mcgee09}, it should in principle be possible to produce the observed falling galaxy number per unit mass with mass as long as the galaxy number density per unit mass is lower in the field \citep[see also discussion in][]{chiu16a}.  At $z\sim1$, indications are that the the stellar mass fraction in the field is lower than in massive galaxy clusters \citep[][]{vdB13,chiu16a}.  If, as we expect, the galaxy number per unit mass is tracked well by the stellar mass fraction, then this suggests that a mix of accretion from the field and from subclusters results in the observed weak or absent redshift trend in $N_{200}$.  Further study of this phenomena using galaxy formation simulations with enough volume to track galaxy populations in cluster mass halos is warranted.

It has been suggested that the growth of the central dominant galaxy within clusters should lead to an observed reduction in the concentration of the galaxies over cosmic time as galaxies merge with the central galaxy \citep{vanderburg13}.  However, our SZE selected sample of massive clusters does not provide statistically significant evidence for a systematic variation of the concentration with redshift in either the full or RS populations.  Moreover, there is no statistically significant trend in the number of galaxies $N_{200}$ within the virial region as a function of redshift (at fixed mass) for the full population.  Thus, the picture for how the central galaxies are built up from the rest of the galaxy population is not yet fully clear. Perhaps with a larger sample we will be able to better discern these changes and measure the impact of the central galaxy growth on the full galaxy population within the virial region.
 
Overall, our study underscores the power of combining a large mm-wave survey from SPT that enables SZE cluster selection with the deep, multi band optical survey dataset from DES.  The selection of the sample is homogeneous and does not directly depend on properties of the galaxy population.  Moreover, each cluster has a high quality SZE mass proxy that has been calibrated to mass over the full redshift range \citep{bocquet15}.   This, together with the deep and wide area DES data, allows us to study the galaxy populations present in the same portion of the virial region in massive galaxy clusters over the last $\sim$10~Gyr period in cosmic evolution.  This initial examination of the galaxy populations within SPT selected clusters will benefit from expansion to the larger sample available today and from an increased focus on the transition of the population from the field to the cluster.

\section*{Acknowledgements}

We acknowledge the support by the DFG Cluster of Excellence ``Origin and Structure of the Universe'', the Transregio program TR33 ``The Dark Universe'' and the Ludwig-Maximilians-Universit\"at. The data processing has been carried out on the computing facilities of the Computational Center for Particle and Astrophysics (C2PAP), located at the Leibniz Supercomputer Center (LRZ).  

The South Pole Telescope is supported by the National Science Foundation through grant PLR-1248097. Partial support is also provided by the NSF Physics Frontier Center grant PHY-1125897 to the Kavli Institute of Cosmological Physics at the University of Chicago, the Kavli Foundation and the Gordon and Betty Moore Foundation grant GBMF 947.  

This paper has gone through internal review by the DES collaboration.  Funding for the DES Projects has been provided by the U.S. Department of Energy, the U.S. National Science Foundation, the Ministry of Science and Education of Spain, the Science and Technology Facilities Council of the United Kingdom, the Higher Education Funding Council for England, the National Center for Supercomputing  Applications at the University of Illinois at Urbana-Champaign, the Kavli Institute of Cosmological Physics at the University of Chicago, the Center for Cosmology and Astro-Particle Physics at the Ohio State University, the Mitchell Institute for Fundamental Physics and Astronomy at Texas A\&M University, Financiadora de Estudos e Projetos, Funda{\c c}{\~a}o Carlos Chagas Filho de Amparo {\`a} Pesquisa do Estado do Rio de Janeiro, Conselho Nacional de Desenvolvimento Cient{\'i}fico e Tecnol{\'o}gico and the Minist{\'e}rio da Ci{\^e}ncia, Tecnologia e Inova{\c c}{\~a}o, the Deutsche Forschungsgemeinschaft and the Collaborating Institutions in the Dark Energy Survey. 

The Collaborating Institutions are Argonne National Laboratory, the University of California at Santa Cruz, the University of Cambridge, Centro de Investigaciones Energ{\'e}ticas, Medioambientales y Tecnol{\'o}gicas-Madrid, the University of Chicago, University College London, the DES-Brazil Consortium, the University of Edinburgh, the Eidgen{\"o}ssische Technische Hochschule (ETH) Z{\"u}rich, Fermi National Accelerator Laboratory, the University of Illinois at Urbana-Champaign, the Institut de Ci{\`e}ncies de l'Espai (IEEC/CSIC), the Institut de F{\'i}sica d'Altes Energies, Lawrence Berkeley National Laboratory, the Ludwig-Maximilians Universit{\"a}t M{\"u}nchen and the associated Excellence Cluster Universe, the University of Michigan, the National Optical Astronomy Observatory, the University of Nottingham, The Ohio State University, the University of Pennsylvania, the University of Portsmouth, SLAC National Accelerator Laboratory, Stanford University, the University of Sussex, Texas A\&M University, and the OzDES Membership Consortium.

The DES data management system is supported by the National Science Foundation under Grant Number AST-1138766. The DES participants from Spanish institutions are partially supported by MINECO under grants AYA2012-39559, ESP2013-48274, FPA2013-47986, and Centro de Excelencia Severo Ochoa SEV-2012-0234.  Research leading to these results has received funding from the European Research Council under the European Union's Seventh Framework Programme (FP7/2007-2013) including ERC grant agreements  240672, 291329, and 306478.

\section*{Affiliations}
\LMU Faculty of Physics, Ludwig-Maximilians-Universit\"at, Scheinerstr. 1, 81679 Munich, Germany \\
\ECU Excellence Cluster Universe, Boltzmannstr.\ 2, 85748 Garching, Germany \\
\MPE Max Planck Institute for Extraterrestrial Physics, Giessenbachstrasse, 85748 Garching, Germany \\
\CTIO Cerro Tololo Inter-American Observatory, National Optical Astronomy Observatory, Casilla 603, La Serena, Chile \\
\UCL Department of Physics \& Astronomy, University College London, Gower Street, London, WC1E 6BT, UK \\
\Rhodes Department of Physics and Electronics, Rhodes University, PO Box 94, Grahamstown, 6140, South Africa \\
\Colby Department of Physics \& Astronomy, Colby College, 5800 Mayflower Hill, Waterville, Maine 04901 \\
\Harvard Department of Physics, Harvard University, 17 Oxford Street, Cambridge, MA 02138, USA \\
\CNRS CNRS, UMR 7095, Institut d'Astrophysique de Paris, F-75014, Paris, France \\
\Sorbonne Sorbonne Universit\'es, UPMC Univ Paris 06, UMR 7095, Institut d'Astrophysique de Paris, F-75014, Paris, France \\
\Carnegie Carnegie Observatories, 813 Santa Barbara St., Pasadena, CA 91101, USA \\
\ICG Institute of Cosmology \& Gravitation, University of Portsmouth, Portsmouth, PO1 3FX, UK \\
\LINEA Laborat\'orio Interinstitucional de e-Astronomia - LIneA, Rua Gal. Jos\'e Cristino 77, Rio de Janeiro, RJ - 20921-400, Brazil \\
\ONB Observat\'orio Nacional, Rua Gal. Jos\'e Cristino 77, Rio de Janeiro, RJ - 20921-400, Brazil \\
\UIA Department of Astronomy, University of Illinois, 1002 W. Green Street, Urbana, IL 61801, USA \\
\NCSA National Center for Supercomputing Applications, 1205 West Clark St., Urbana, IL 61801, USA \\
\ICE Institut de Ci\`encies de l'Espai, IEEC-CSIC, Campus UAB, Carrer de Can Magrans, s/n,  08193 Bellaterra, Barcelona, Spain \\
\IFAE Institut de F\'{\i}sica d'Altes Energies (IFAE), The Barcelona Institute of Science and Technology, Campus UAB, 08193 Bellaterra (Barcelona) Spain \\
\Southhampton School of Physics and Astronomy, University of Southampton,  Southampton, SO17 1BJ, UK \\
\FNAL Fermi National Accelerator Laboratory, P. O. Box 500, Batavia, IL 60510, USA \\
\UPPA Department of Physics and Astronomy, University of Pennsylvania, Philadelphia, PA 19104, USA \\
\JPL Jet Propulsion Laboratory, California Institute of Technology, 4800 Oak Grove Dr., Pasadena, CA 91109, USA \\
\UMA Department of Astronomy, University of Michigan, Ann Arbor, MI 48109, USA \\
\UMP Department of Physics, University of Michigan, Ann Arbor, MI 48109, USA \\
\KavliC Kavli Institute for Cosmological Physics, University of Chicago, Chicago, IL 60637, USA \\
\UFA Department of Astronomy, University of Florida, Gainesville, FL 32611 \\
\KavliS Kavli Institute for Particle Astrophysics \& Cosmology, P. O. Box 2450, Stanford University, Stanford, CA 94305, USA \\
\SLAC SLAC National Accelerator Laboratory, Menlo Park, CA 94025, USA \\
\CAPP Center for Cosmology and Astro-Particle Physics, The Ohio State University, Columbus, OH 43210, USA \\
\OSUP Department of Physics, The Ohio State University, Columbus, OH 43210, USA \\
\Montreal D\'epartement de Physique, Universit\'e de Montr\'eal, C.P. 6128, Succ. Centre-Ville, Montr\'eal, Qu\'ebec H3C 3J7, Canada \\
\AAO Australian Astronomical Observatory, North Ryde, NSW 2113, Australia \\
\Texas George P. and Cynthia Woods Mitchell Institute for Fundamental Physics and Astronomy, and Department of Physics and Astronomy, Texas A\&M University, College Station, TX 77843,  USA \\
\OSUA Department of Astronomy, The Ohio State University, Columbus, OH 43210, USA \\
\KavliM Kavli Institute for Astrophysics and Space Research, Massachusetts Institute of Technology, 77 Massachusetts Avenue, Cambridge, MA 02139 \\
\Princeton Department of Astrophysical Sciences, Princeton University, Peyton Hall, Princeton, NJ 08544, USA \\
\ICRE Instituci\'o Catalana de Recerca i Estudis Avan\c{c}ats, E-08010 Barcelona, Spain \\
\UMelb School of Physics, University of Melbourne, Parkville, VIC 3010, Australia \\
\Sussex Department of Physics and Astronomy, Pevensey Building, University of Sussex, Brighton, BN1 9QH, UK \\
\UAP Department of Physics, University of Arizona, Tucson, AZ 85721, USA \\
\CIE Centro de Investigaciones Energ\'eticas, Medioambientales y Tecnol\'ogicas (CIEMAT), Madrid, Spain \\
\IFB Instituto de F\'\i sica, UFRGS, Caixa Postal 15051, Porto Alegre, RS - 91501-970, Brazil \\
\UAH Institute for Astronomy, University of Hawaii at Manoa, Honolulu, HI 96822, USA \\
\CFA Harvard-Smithsonian Center for Astrophysics, 60 Garden Street, Cambridge, MA 02138 \\
\UCD Physics Department, University of California, Davis, CA 95616 \\
\ANL Argonne National Laboratory, 9700 South Cass Avenue, Lemont, IL 60439, USA \\



\bibliographystyle{mnras}
\bibliography{paper} 

\bsp	
\label{lastpage}
\end{document}